\documentclass[prd,aps,eqsecnum,amsmath,floatfix,nofootinbib,preprint,tightenlines]{revtex4}

\usepackage{latexsym}
\usepackage{graphicx}
\usepackage{multirow}
\usepackage[dvipsnames]{xcolor}
\usepackage{hyperref}
\def\NON{\nonumber\\}
\def\bibi{\bibitem}

\def\a{\alpha}
\def\b{\beta}
\def\c{\chi}
\def\d{\delta}
\def\e{\epsilon}                % Also, \varepsilon
\def\g{\gamma}

\def\j{\psi}

\def\l{\lambda}
\def\m{\mu}
\def\n{\nu}

\def\p{\pi}                     % Also, \varpi
\def\th{\theta}                  %       \vartheta
\def\r{\rho}                    %       \varrho
\def\s{\sigma}                  %       \varsigma
\def\t{\tau}

\def\z{\zeta}
\def\D{\Delta}

\def\G{\Gamma}

\def\L{\Lambda}
\def\O{\Omega}

\def\S{\Sigma}
\def\U{\Upsilon}

\def\ca{{\cal A}}
\def\cb{{\cal B}}

\def\cl{{\cal L}}
\def\cm{{\cal M}}

\def\cp{{\cal P}}

\def\cs{{\cal S}}

\def\cv{{\cal V}}

\def\bo{\raisebox{-.4ex}{\large$\Box$}}                 % D'Alembertian
\def\cbo{{\,\raise-.15ex\Sc [\,}}                       % curly "
\def\ltap{\raisebox{-.4ex}{\rlap{$\sim$}} \raisebox{.4ex}{$<$}}   % < or ~
\def\Sl#1{\rlap{\hbox{$\mskip 3 mu /$}}#1}      % " upper
\def\bra#1{\Big\langle #1\Big|}                 % < |
\def\ket#1{\Big| #1\Big\rangle}                 % | >
\def\vev#1{\Big\langle #1 \Big\rangle}           % < >
\def\svev#1{\left\langle #1\right\rangle}       % variable < >
\def\ddt#1{{\buildrel {\hbox{\LARGE .\kern-2pt.}} \over {#1}}}% double dot-over
\def\ie{\mbox{\it i.e.}}
\def\eg{\mbox{\it e.g.}}

\def\leqx{\,\raisebox{-1.0ex}{$\stackrel{\textstyle <}{\sim}$}\,}

\def\tr{{\rm tr}\,}

\def\half{{1\over 2}}

\def\ttl#1{{\it #1}}
\def\tr{\,{\rm tr}}
\def\vev#1{\langle #1\rangle}
\def\bra#1{\langle #1|}
\def\ket#1{|#1\rangle}

\def\tg{\tilde{g}}

\def\tQ{\tilde{Q}}
\def\tilt{\tilde\t}

\def\ta{\tilde\a}
\def\tb{\tilde\b}
\def\tG{\tilde\G}

\def\tO{\tilde\O}

\def\tcl{\tilde\cl}

\def\hB{\hat{B}}
\def\hatc{\hat{c}}

\def\hf{\hat{f}}

\def\bj{\overline{\j}}

\def\Th{\Theta}

\def\tc{\tilde{c}}

\def\GMIC{\G^{\rm MIC}}

\def\GEFT{\G^{\rm EFT}}
\def\tGEFT{\tG^{\rm EFT}}

\begin{document}

\begin{center}
{\large\bf Low-energy effective action for pions and a dilatonic meson}\\[8mm]
Maarten Golterman$^a$ and Yigal Shamir$^b$\\[8 mm]
{\small
$^a$Department of Physics and Astronomy, San Francisco State University,\\
San Francisco, CA 94132, USA\\
$^b$Raymond and Beverly Sackler School of Physics and Astronomy,\\
Tel~Aviv University, 69978, Tel~Aviv, Israel}\\[10mm]
\end{center}

\begin{quotation}
Numerical simulations of QCD-like theories in which the number of flavors
is adjusted so that the beta function is very small, but confinement
and chiral symmetry breaking nevertheless take place, appear to
reveal the presence
of a flavor-singlet scalar meson which can be as light as the pions.
Because the breaking of dilatation symmetry, quantified by the beta function,
is small relative to QCD, a possible explanation is that the scalar meson
is a pseudo Nambu-Goldstone boson associated with the approximate
dilatation symmetry.  We use this observation to systematically develop
a low-energy effective action that accounts for both the pions and
the ``dilatonic'' scalar meson.  In order to justify the power counting
that controls the couplings of the dilatonic meson we invoke the Veneziano
limit, in which the number of fundamental-representation flavors $N_f$
grows in proportion with the number of colors $N_c$, while the ratio $N_f/N_c$
is kept close to, but below, the critical value where the conformal window
is entered.
\end{quotation}

%%%%%%%%%%%%%%%%%%%%%%%%%%%
\newpage
\section{\label{intro} Introduction}
%%%%%%%%%%%%%%%%%%%%%%%%%%%

A non-abelian gauge theory with $N_c$ colors and $N_f$ flavors
of massless Dirac fermions can be in one of three phases. For
small $N_f$, the renormalized coupling becomes strong toward the
infrared, leading to confinement and chiral symmetry breaking.
As $N_f$ is increased the running slows down until eventually
an infrared-attractive fixed point (IRFP) develops \cite{WEC,BZ}.
The long-distance theory is then conformal,
exhibiting neither confinement nor chiral symmetry breaking.
Finally, if $N_f$ gets too large, asymptotic freedom is lost.

In this paper we consider ``walking'' theories, where confinement and
chiral symmetry breaking still occur, but $N_f$ has become so large
that the theory is on the verge of developing an IRFP.  This is manifested
in a non-perturbative beta function that is much smaller than in QCD.\footnote{
  For a recent review of lattice results, see Ref.~\cite{TD}.
}
Recent lattice studies of some candidate walking theories
reveal the presence of a flavor-singlet,
parity-even scalar meson that can be as light as the pions
at the fermion masses used in those simulations, and much lighter
than any other hadronic states in the theory.
A light scalar meson was found in the $SU(3)$ theory with $N_f=8$
Dirac fermions in the fundamental representation \cite{LatKMIs,LSDs},
which current evidence favors to be chirally broken
and confining \cite{LatKMIc,LSD,CHPS,HSV,BMW8}.
A light scalar was also found in the $SU(3)$ theory with two Dirac fermions
in the sextet (two-index symmetric) representation \cite{BMWs}.
Because the non-perturbative study of the beta function
is very difficult \cite{TACO},
it is still not entirely certain
whether this model is chirally broken or infrared conformal.
While the results of Ref.~\cite{BMW6} favor the chirally broken option,
other recent studies are inconclusive \cite{HLH,JKDS}.

The standard low-energy effective theory for the pions---the pseudo
Nambu-Goldstone bosons (pNGBs) of spontaneously broken chiral symmetry---is
the chiral lagrangian \cite{SW,GL1,GL2}.\footnote{
  For a recent review of chiral perturbation theory with applications
  to lattice QCD, see Ref.~\cite{MGrev}.
}
But when the scalar meson is as light as the pions, a consistent, unitary
low-energy effective theory must accommodate this scalar state as well.
Moreover, if the scalar is much lighter than all other states apart from
the pions, it is
natural to ask for an effective field theory description of this scalar,
even if fermion masses are taken so small that the pions become lighter
than this scalar.

A natural question is whether, like the pions, the light scalar meson
can be interpreted as a pNGB associated with the spontaneous breaking
of another approximate symmetry.  In this paper we will attempt to provide
such a framework, where the new approximate symmetry is the
invariance under dilatations, or scale transformations.
For this reason, we will refer to the scalar meson
in this paper as the dilatonic meson.\footnote{%
  For a previous treatment of this problem, see Ref.~\cite{CT}.
  For a discussion of the effective theory for a light dilatonic scalar
  in theories with approximate scale invariance, but without
  spontaneously broken of chiral symmetry, see Ref.~\cite{ChM}.
}

It is far from obvious that such an interpretation makes sense.
In the case of chiral symmetry, the fermion mass $m$ is a parameter that
can be tuned continuously such that, for $m\to 0$, exact chiral symmetry
is recovered, and the pions become massless.
By contrast, while scale invariance is broken softly by the
fermion mass, its breaking by the scale dependence
of the renormalized coupling constitutes a hard breaking.
Even if we dial the fermion mass to zero,
we do not have another parameter that would allow us to continuously control
the explicit breaking of scale invariance encoded in the beta function.
Moreover, if ultimately confinement and chiral symmetry breaking are to
take place,
it is unavoidable that the coupling will keep growing as the energy scale
is lowered.  The hard breaking of dilatation symmetry by the running
(or walking) of the coupling is therefore of a very different nature than
the soft breaking of chiral symmetry by a fermion mass term.

A hint is provided by the effective low-energy treatment of the anomalous
axial symmetry $U(1)_A$.  In this case, too, there is no way to ``turn off''
the anomaly within a given theory.  In order to overcome this problem
one takes the familiar large-$N_c$ limit: the number of colors $N_c$ is taken
to infinity, while the number of flavors  is held fixed
\cite{Hooft,Witten,VV,largeNmore}.
In this limit the axial anomaly becomes vanishingly small,
and the flavor-singlet pseudoscalar meson
eventually becomes as light as the pions.  A large-$N_c$ extension
of the standard chiral lagrangian that accommodates the $U(1)_A$ symmetry
was systematically developed in Refs.~\cite{GL2,KL}.

Here we will invoke a different large-$N$ limit \cite{VZlimit}.
Considering $SU(N_c)$ gauge theories with $N_f$ Dirac fermions
in the fundamental representation, the so-called Veneziano limit
is defined as the limit $N\to\infty$, in which we take $N_c=N$,
while the ratio
\begin{equation}
  n_f = \frac{N_f}{N_c} \ ,
\label{nf}
\end{equation}
is held fixed.  Asymptotic freedom imposes the restriction $n_f < 11/2$.
For each $N_c$, we let $N_f^*(N_c)$ be the largest number
of flavors such that the theory confines, and chiral symmetry is
broken spontaneously in the chiral limit $m\to 0$.
The {\it conformal window}, in which the coupling runs into an IRFP
and the long-distance theory is conformal, thus corresponds to $N_f$ values
in the range
\begin{equation}
  N_f^*(N_c) < N_f < (11/2) N_c \ .
\label{cwindow}
\end{equation}
In the Veneziano limit, the conformal window becomes
\begin{equation}
   n_f^* < n_f < 11/2 \ ,
\label{cwinVlim}
\end{equation}
where
\begin{equation}
  n_f^* = \lim_{N_c\to\infty} \frac{N_f^*(N_c)}{N_c} \ .
\label{nfstar}
\end{equation}

In this paper we are interested in chirally broken theories,
and thus we will only consider theories for which
\begin{equation}
  n_f < n_f^* \ .
\label{nfdiff}
\end{equation}
Since $N_c$ and $N_f$ are integers, the increments one can make in $n_f$
are steps of size $1/N$.  It follows that, in the Veneziano limit,
$n_f$ effectively becomes a continuous parameter.
For the fundamental representation, the increments in the coefficients of
the two-loop beta function also come in steps of size $1/N$,
suggesting that $n_f - n_f^*$ is the parameter controlling the smallness
of the beta function, or the ``walking'' nature of the theory.

Our analysis will be based on the
conjecture that dilatation symmetry is recovered in a {\it triple limit}:
the chiral limit $m\to 0$; the Veneziano large-$N$ limit; and finally
the limit $n_f\nearrow n_f^*$.  We will assume that in this triple limit
the trace anomaly becomes vanishingly small towards the infrared,
both inside and outside of the conformal window.  Correspondingly,
the effective low-energy theory developed in this paper aims to provide
the broken phase with
a systematic expansion in three small parameters: the fermion mass,
$1/N$, and $n_f - n_f^*$.  As usual, it involves a derivative expansion
as well.

\medskip

This paper is organized as follows.  In Sec.~\ref{mic} we discuss the
microscopic theory.  We briefly review the trace anomaly,
and show how it can be encoded in the coupling of the microscopic theory
to an external dilaton source $\s(x)$.  In addition, we introduce
the familiar external sources for scalar and pseudoscalar fermion bilinears.
As usual, the external fields allow us to match correlation functions between
the microscopic and the effective theory, and ultimately, to determine
the coupling constants of the effective theory order by order
in the low-energy expansion.

In Sec.~\ref{LOV} we begin with a more detailed discussion of the triple limit
which is the starting point of our low-energy expansion.
We specify the assumptions, or conjectures,
that must be made in order to justify the expansion.

Having introduced the power counting, we proceed to address
a complication in the construction of the low-energy theory
that was previously encountered in the treatment of the $U(1)_A$
symmetry in large $N_c$ \cite{GL2,KL}.  In principle,
each term in the effective low-energy lagrangian can be multiplied
by an arbitrary potential $V(\t-\s)$, where $\t(x)$ is
the effective field that represents the dilatonic meson.
Like the $U(1)_A$ symmetry, the dilatation symmetry is abelian.
As a result, it does not impose any algebraic constraints on these potentials,
and the coefficients of their power-series expansion in $\t-\s$
are independent parameters of the low-energy theory.
If nothing further could be done, the power-series coefficients
would constitute an infinitely large set of low-energy parameters,
and our effective field theory ends up having no predictive power.
Fortunately, however, the matching of
the microscopic and the effective theories for $n_f\nearrow n_f^*$
imposes a power-counting
hierarchy on the coefficients of the potentials.
As a consequence, out of the four potentials
that we encounter {\it a-priori} in the leading-order lagrangian,
three are eliminated altogether.  The remaining potential,
denoted $V_d$, gets truncated to a linear function
\begin{equation}
  V_d(\t-\s) = c_0 + c_1 (\t-\s) \ .
\label{Vd01}
\end{equation}

In Sec.~\ref{LO} we proceed to study the classical ground state,
which turns out to have some delicate features.
In the chiral limit, the classical vacuum is controlled by the
potential of the dilatonic meson, which in turn
is proportional to $e^{4\t} V_d(\t)$.
The parameter $c_1$, introduced in Eq.~(\ref{Vd01}) above,
is the only parameter of the leading-order lagrangian
that accounts for the hard breaking of scale invariance
via the walking of the coupling in the microscopic theory.
We find that the $c_1$ term is crucial for the existence
of a stable classical vacuum.  We also study the tree-level masses,
the dilatation current, and the fermion condensate.

As an application, we study in Sec.~\ref{NLO} the one-loop effective potential.
This provides a concrete example that the renormalization of the
effective theory works in essentially the same way as in standard
chiral perturbation theory.  We summarize our results in Sec.~\ref{disc}.
Various technical details are relegated to several appendices.

%%%%%%%%%%%%%%%%%%%%%%%%%%%
%\newpage
\section{\label{mic} The microscopic theory}
%%%%%%%%%%%%%%%%%%%%%%%%%%%
In this section we discuss the relevant properties of the microscopic theory.
We begin in Sec.~\ref{dST} with a brief review of the dilatation current
and the trace anomaly.  As a further gearing up, we recall
in Sec.~\ref{Chisources} the basic steps in the construction
of the chiral lagrangian.  Apart from the obviously needed dynamical
effective fields, a set of source fields with appropriate
transformation properties, or spurions, is needed to match
the effective theory to the low-energy sector of the microscopic theory.
We also review how the anomalous $U(1)_A$ symmetry of QCD can be accommodated
in the effective field theory framework, since this provides us
with a useful lesson.  In Sec.~\ref{Levnr} we return to dilatation symmetry,
and discuss how the operatorial trace
of the energy-momentum tensor can be encoded
in the coupling of the microscopic theory
to an external source with the transformation properties of a dilaton.

%%%%%%%%
%\newpage
\subsection{\label{dST} Dilatation current and trace anomaly}
%%%%%%%%
The dilatation current is given by
\begin{equation}
  S_\m = x_\n T_{\m\n} \ .
\label{SxT}
\end{equation}
where $T_{\m\n}$ is the symmetric, gauge-invariant energy-momentum tensor.
Being a conserved current, $T_{\m\n}$ does not renormalize.\footnote{%
  This would not be true in the presence of elementary scalar fields,
  because of the existence of an ``improvement term'' with which
  $T_{\m\n}$ can mix \cite{CCJ,JCC}.
}
In the asymptotically-free gauge theories that we will study in this paper
the dilatation current is not conserved.  Its divergence is given
by the trace of the energy-momentum tensor,
\begin{equation}
  \partial_\m S_\m  = T_{\m\m} \equiv -T  \ .
\label{dS}
\end{equation}
On shell, the trace of the energy-momentum tensor is \cite{CDJ}
\begin{subequations}
\label{rentrace}
\begin{eqnarray}
  T &=& T_{cl} + T_{an} \ ,
\label{rentracea}\\
  T_{cl} &=& m[\bj\j] \ ,
\label{rentraceb}\\
  T_{an} &=& \, \tb\, [F^2] + \g_m\, m\, [\bj\j] \ .
\label{rentracec}
\end{eqnarray}
\end{subequations}
The operatorial trace divides into a classical part $T_{cl}$
and a quantum part $T_{an}$.  We use the shorthand $F^2 = F_{\m\n}^a F_{\m\n}^a$,
where $F_{\m\n}^a$ is the non-abelian field strength,
while $[\cdots]$ denotes renormalized operators.
Since $T_{\m\n}$ is renormalization-group invariant, so is $S_\m$,
and thus $T$.  Moreover, $m[\bj\j]$ is invariant by itself,
and hence $T_{cl}$ and $T_{an}$ are separately
renormalization-group invariant.\footnote{%
  For an explicit verification, see Ref.~\cite{RT}.
}
The renormalization-group evolution of the renormalized coupling and
fermion mass is controlled by
\begin{eqnarray}
  \tb &=& \frac{\b(g)}{2g} \ , \hspace{10ex}
  \b(g) \ = \ \m \frac{\partial g}{\partial\m} \ ,
\label{beta}\\
  \g_m &=& \g_m(g) \ = \ -\frac{\m}{m} \frac{\partial m}{\partial\m} \ .
\label{gamma}
\end{eqnarray}

%%%%%%%%
%\newpage
\subsection{\label{Chisources} Chiral symmetry and spurions}
%%%%%%%%
The explicit breaking of dilatation symmetry must be incorporated into
the low-energy effective theory that we will construct
in the following sections.  We can learn how to do this
from the conceptually similar problem of extending the chiral lagrangian
to accommodate the anomalous $U(1)_A$ symmetry in the
't Hooft large-$N_c$ limit.

We first recall the key steps in the construction of
the standard chiral lagrangian \cite{SW,GL1,GL2}.
We introduce a dynamical effective field $\S(x)$ taking values
in the Nambu--Goldstone-boson manifold, and endow it with appropriate
transformation properties under chiral symmetry.  For $N_f$ Dirac fermions
in a complex representation,
the chiral symmetry is $SU(N_f)_L\times SU(N_f)_R$.  The effective field
is unitary, $\S(x)\in SU(N_f)$, and transforms according to
\begin{equation}
  \S(x) \to g_L \S(x) g_R^\dagger \ , \hspace{5ex}
  g_{L,R} \in SU(N_f)_{L,R} \ .
\label{Sigmatrans}
\end{equation}

The order-by-order construction of the effective lagrangian is dictated
by the power counting
\begin{equation}
  \d \sim m/\L_{IR}  \sim p^2/\L_{IR}^2  \ .
\label{chiralpc}
\end{equation}
Here $\d$ is a generic name for the small expansion parameter of the
effective theory, $m$ is the fermion mass,
and $p^2$ is a characteristic momentum(-squared) of the pNGBs.
$\L_{IR}$ is the non-perturbative dynamical scale of the theory
which we will discuss later on (see Sec.~\ref{conj}).
At leading order the effective lagrangian consists
of terms of order $\d$ in the power counting, the next to leading order
consists of terms of order $\d^2$, and so on.

The effective lagrangian at each order is fixed
by matching correlation functions calculated in the microscopic theory
with the corresponding correlation functions calculated in the effective theory.
In order to facilitate this matching, a common set of source fields
is introduced in both the microscopic and effective theories.
In the matching process we can make use of any correlation function
obtained by multiple differentiations of the (microscopic and effective)
partition functions with respect to the external sources.
For the basic chiral lagrangian,
one needs a scalar source $\cs_{AB}(x)$, $A,B=1,2,\ldots,N_f$,
which couples to the bilinear $\bj_A(x)\j_B(x)$;
and a pseudoscalar source $\cp_{AB}(x)$ that
couples to the bilinear $\bj_A(x) i\g_5\j_B(x)$.
The chiral transformation properties are
\begin{subequations}
\label{LRtrans}
\begin{eqnarray}
  \j &\to& (g_R P_R + g_L P_L) \j \ ,
\label{LRtransa}\\
  \bj &\to& \bj (P_L g_R^\dagger + P_R g_L^\dagger) \ ,
\label{LRtransb}\\
  \c &\to& g_L \c\, g_R^\dagger \ ,
\label{LRtransc}
\end{eqnarray}
\end{subequations}
where we have promoted the external sources to spurions, and
\begin{equation}
  \c = \cs+i\cp \ .
\label{LRtransd}
\end{equation}
This allows us to add the invariant
\begin{equation}
  \cl_m=\bj(\c P_R+\c^\dagger P_L)\j \ ,
\label{micmass}
\end{equation}
to the microscopic lagrangian, thus coupling the theory to the sources
$\cs$ and $\cp$.

The actual theory is recovered by setting the source fields
to their ``expectation values,''
\begin{equation}
  \cs(x) = m \ , \qquad \cp(x) = 0 \ .
\label{spurvev}
\end{equation}
For simplicity, we assume in this paper that all fermions have a common
mass $m$.  Thus, chiral symmetry is softly broken for $m\ne0$.
By dialing $m\to 0$ exact chiral symmetry is recovered,
and the pions become massless.

In general, any dependence of the effective theory
on the expectation value of a spurion field encodes some explicit breaking
of a symmetry, because the effective lagrangian is only invariant under
the simultaneous transformation of the dynamical and spurion (source)
fields.  In the standard chiral lagrangian, this manifests itself
as the breaking of chiral symmetry by the fermion mass,
as we have just discussed.
Let us next see how this comes into play when dealing with
the anomalous $U(1)_A$ symmetry.

The axial anomaly can be formally cancelled by
augmenting the lagrangian of the microscopic theory with a new term
\begin{equation}
  -c g^2 N_f\, \th(x) \tr F(x)\tilde{F}(x) \ .
\label{FFdual}
\end{equation}
Here $\th(x)$ is the axion field, which couples to (minus) the anomalous
divergence of the singlet axial current.  Under a $U(1)_A$ transformation
the axion field transforms by a shift, and this allows us to cancel
the anomaly generated by the dynamical fermions against the variation
of the axion-dependent term, Eq.~(\ref{FFdual}).

The low-energy theory will now depend on the axion field $\th(x)$.
As long as we treat the axion field as a spurion, which transforms
under the $U(1)_A$ symmetry, the effective lagrangian will maintain $U(1)_A$
invariance.  But once we assign to the axion field a fixed expectation value,
the axial anomaly will manifest itself in the effective theory.

The expectation value of the axion field is the vacuum angle $\th_0$.
The effective theory fails to be invariant under $U(1)_A$
even if we set the vacuum angle to zero.
The effective theory now includes an additional dynamical field $\eta(x)$
associated with the flavor-singlet pseudoscalar meson,
which couples to the $U(1)_A$ current.  Algebraically, the $\eta(x)$ field
is similar to the axion, in that they both transform by a shift under $U(1)_A$.
Any function of the difference
$\eta-\th$ is invariant under the abelian $U(1)_A$ transformations.
As discussed in Refs.~\cite{GL2,KL}, already at leading order
the effective low-energy theory depends on several
such arbitrary non-derivative functions, or potentials
$V_I(\eta-\th)$.\footnote{%
  Because of parity, $V_I(\eta-\th)$ is an even function of its argument.
}
When we set $\th(x)=\th_0=0$, all that happens is that $V_I(\eta-\th)$
becomes $V_I(\eta)$.  The presence in the
effective theory of the $V_I(\eta)$ potentials represents
an explicit breaking of $U(1)_A$.  The question of how this arbitrariness
can be controlled is postponed to Sec.~\ref{LOV}.

%%%%%%%%
%\newpage
\subsection{\label{Levnr} Encoding the explicit breaking of dilatations}
%%%%%%%%
As explained in the previous subsection,
the construction of the low-energy effective theory begins with
the identification of the dynamical effective fields
and the external source fields.  The latter are then promoted to spurions.
The new effective field that describes the dilatonic meson
will be introduced in the next section.  Here we discuss how to encode
the anomalous divergence of the dilatation current by coupling
the microscopic theory to an external dilaton field $\s(x)$.

We consider an asymptotically free gauge theory with $N_f$ Dirac fermions
in the fundamental representation.
After a rescaling of the bare gauge and fermion fields by the
bare coupling, we express the bare action in $d$ dimensions as
\begin{eqnarray}
  S_{ev}(\s) &=& \int d^dx\, e^{(d-4)\s(x)} \cl_0(x) \ ,
\label{Sbare}\\
  \cl_0 &=& \frac{1}{g_0^2} ( \cl_k + \cl_{src} ) \ ,
\label{bareL}\\
  \cl_k &=& \frac{1}{4} F_{\m\n}^a F_{\m\n}^a
  + \sum_{i=1}^{N_f} \bj_i \Sl{D} \j_i \ ,
\label{bareL0}\\
  \cl_{src} &=& \bj(\cs + i\cp\g_5)\j \ ,
\label{bareSP}
\end{eqnarray}
where the covariant derivative is
$D_\m = \partial_\m + iA_\m$, and the field strength is
$F_{\m\n} = F_{\m\n}^a T^a = -i[D_\m,D_\n]$.

The dilaton field $\s(x)$ has been introduced in Eq.~(\ref{Sbare}) via
an overall factor multiplying the entire bare lagrangian.
We have denoted the action by $S_{ev}(\s)$, with
a subscript ``$ev$'' to indicate that the
dependence on $\s(x)$ comes from the evanescent part of the bare lagrangian.
$S_{ev}(\s)$ is manifestly invariant under the $d$-dimensional dilatation
\begin{subequations}
\label{baredil}
\begin{eqnarray}
  A_\m(x) &\to& \l A_\m(\l x) \ ,
\label{baredila}\\
  \j(x) &\to& \l^{3/2} \j(\l x) \ , \qquad
  \bj(x) \ \to \ \l^{3/2} \bj(\l x) \ ,
\label{baredilb}\\
  \cs(x) &\to& \l \cs(\l x) \ , \hspace{7ex}
  \cp(x) \ \to \ \l \cp(\l x) \ ,
\label{baredilc}\\
  \s(x) &\to& \s(\l x) + \log \l \ ,
\label{baredild}\\
  g_0 &\to& g_0 \ .
\label{baredile}
\end{eqnarray}
\end{subequations}
With the obvious exception of $\s(x)$, which plays a special role,
in this setup the
(dynamical and external) bare fields transform according to their canonical
mass dimension in $d=4$ dimensions.  The bare parameter $g_0$ is dimensionful
in $d\ne 4$ dimensions, but the anticipated transformation of $1/g_0^2$
has been taken up by the factor of $e^{(d-4)\s(x)}$.

In order to see how the $\s$ dependence encodes the anomalous trace
of the energy-momentum tensor, $T_{an}$, we proceed to consider the renormalized
lagrangian.  Differentiating the bare action once with respect to
$\s(x)$ and then setting it to zero we have
\begin{equation}
  \frac{\partial}{\partial \s(x)}\, S_{ev}(\s) \bigg|_{\s(x)=0}
  = (d-4) \cl_0(x) \ .
\label{dSdsig}
\end{equation}
It was shown in Ref.~\cite{CDJ} that the evanescent operator on the
right-hand side occurs in the $d$-dimensional trace of the energy-momentum
tensor, alongside with the classical (four-dimensional) trace,
and with terms proportional to equations of motion.
Taking the $d\to 4$ limit and discarding terms proportional to equations
of motion then yields $T_{an}(x)$.  Hence
\begin{eqnarray}
  S_{ev}(\s) &=& S_{ren}(\s) + O(\s^2) \ ,
\label{evnra}\\
  S_{ren}(\s) &=& \int d^dx\, \Big( \cl_k^{ren} + T_{cl}(\cs,\cp)
  + \s\, T_{an}(\cs,\cp) \Big) \ ,
\label{evnrb}
\end{eqnarray}
where $\cl_k^{ren}$ is equal to $g_0^{-2}\cl_k$ reexpressed
in terms of the renormalized fields
and coupling, and the renormalization scale $\m$.
Here (compare Eq.~(\ref{rentrace}))
\begin{subequations}
\label{traceSP}
\begin{eqnarray}
  T_{cl}(\cs,\cp) &=& [\bj \cs \j] + [\bj\, i\cp \g_5 \j] \ ,
\label{traceSPa}\\
  T_{an}(\cs,\cp) &=& \tb\, [F^2]
  + \g_m\, \Big( [\bj \cs \j] + [\bj\, i\cp \g_5 \j] \Big) \ .
\label{traceSPb}
\end{eqnarray}
\end{subequations}
Chiral symmetry constrains the renormalization of the sources
$\cs(x)$ and $\cp(x)$ to be the same as that of the fermion mass $m$.
The operators $T_{cl}(\cs,\cp)$ and $T_{an}(\cs,\cp)$ are therefore
renormalization-group invariant, too.

The $O(\s^2)$ terms in Eq.~(\ref{evnra}) are needed to ensure
exact dilatation invariance of the regularized action in $d$ dimensions.
However, our derivation of the low-energy action will not depend
on these terms, and therefore we do not consider them any further.

%%%%%%%%%%%%%%%%%%%%%%%%%%%
%\newpage
\section{\label{LOV} Power counting and the leading-order lagrangian}
%%%%%%%%%%%%%%%%%%%%%%%%%%%
In QCD, exact chiral symmetry is restored when the fermion mass $m$
tends to zero.  The chiral lagrangian is a low-energy expansion
around the chiral limit.  The systematic nature of the low-energy
expansion derives from the fact that $m$ is a parameter of the
microscopic theory that can be varied continuously.
When the fermion mass vanishes, so does the mass of the pions.

As mentioned in the introduction, the anomalous $U(1)_A$ symmetry,
and its associated flavor-singlet pseudoscalar meson,
can be incorporated into the framework of the low-energy expansion
by invoking the 't Hooft large-$N_c$ limit, in which  the number of colors $N_c$
tends to infinity, while the number of flavors is held fixed.
For $m=0$, the anomalous divergence of the singlet axial current is
\begin{equation}
   \partial_\m J_{5\m} = cg^2 N_f \tr F\tilde{F} \ ,
\label{dJ5}
\end{equation}
where the numerical constant $c$ depends on the representation
in which the fermions transform.
The axial anomaly is proportional to $g^2$, which scales as $1/N_c$,
because in large-$N_c$ counting the product $g^2 N_c$ is $O(1)$.  As a result,
for $N_c\to\infty$ (and fixed $N_f$)
the mass of the flavor-singlet pseudoscalar meson
tends to zero.\footnote{%
  The actual argument involves a comparison of the large-$N_c$
  scaling of $\vev{ J_{5\m}(x) J_5(y)}$ and of $\vev{\tr F\tilde{F}(x) J_5(y)}$
  where $J_5=\bj\g_5\j$.
}

The starting point of the systematic low-energy expansion
then involves a double limit: the chiral limit $m\to 0$,
together with the limit $N_c\to\infty$.
The power-counting suppression of the axial anomaly in the microscopic theory
is communicated to the low-energy effective theory
by matching correlation functions obtained by differentiating both
partition functions with respect to the axion source $\th(x)$.
Using this, one can show that the coefficients of the power-series
expansion of the $V_I(\eta-\th)$ potentials, introduced in Sec.~\ref{Chisources},
adhere to a natural power-counting hierarchy.
This limits the number of linearly independent low-energy constants
occurring at each order of the expansion to be finite \cite{GL2,KL}.

In our case, we are interested in the different limit already introduced
in Sec.~\ref{intro}, in which the dilatation current is effectively conserved
in the infrared, while the axial current of Eq.~(\ref{dJ5}) remains anomalous.
In Sec.~\ref{conj} we will discuss
this triple limit and the assumptions on which it depends in more detail,
and use it to control the low-energy expansion.
We then proceed in Sec.~\ref{match}
to write down the leading-order lagrangian of the effective theory.
Much like in the case of the $U(1)_A$ symmetry,
using only symmetry considerations and the usual chiral power counting
allows the leading-order lagrangian to depend on four arbitrary potentials
$V_I(\t-\s)$.  Using a strategy similar to that followed in Refs.~\cite{GL2,KL},
we then match correlation functions obtained by
differentiating with respect to the dilaton source $\s(x)$ between
the microscopic and the effective theories, and use this to
establish a power-counting hierarchy for the power-series coefficients
of these potentials.

The result is a leading-order lagrangian
that depends only on a finite (and rather small) number of low-energy
parameters.  In Sec.~\ref{mutrans} we show that dilatation invariance can be maintained even after we set $\s(x)=0$,
at the price of promoting the power-series coefficients of the $V_I(\t)$
potentials to (global) spurions.  We also determine the transformation
properties of the renormalized chiral sources, which fixes the dependence
of the leading-order lagrangian on the mass anomalous dimension.

%%%%%%%%%%%%%%%%%%%%%%%%%%%
%\newpage
\subsection{\label{conj} Nature of the expansion}
%%%%%%%%%%%%%%%%%%%%%%%%%%%
As explained in the introduction, we conjecture that the effective low-energy
theory can be constructed as a systematic expansion around a triple limit.
One of these limits is just the chiral limit, $m\to 0$.
The other two limits, namely, the Veneziano large-$N$ limit
and the limit $n_f\nearrow n_f^*$, are more subtle.

First, we need to define a reference scale against which to measure
dimensionful quantities in the effective theory.
We will set the reference infrared scale using one of the decay constants,
the pion decay constant $\hf_\p$ or the dilatonic-meson decay constant $\hf_\t$,
in the chiral limit.  The decay constants are defined by\footnote{%
  To avoid confusion, we note that Eqs.~(\ref{fpidef}) and~(\ref{ftaudef}) are used
  to define the decay constants for any $m$.  However, we will reserve
  the notation $\hf_\p,\hf_\t$ for the decay constants in the chiral limit.
}
\begin{eqnarray}
 \bra{0} J_{5\m}^a(x) \ket{\pi^b}
  &=& ip_\m \hf_\p\, \d^{ab}\,e^{ipx}\, \ ,
\label{fpidef}\\
  \bra{0} T_{\m\n}(x) \ket{\t}
  &=& \frac{\hf_\t}{3} (-\d_{\m\n} p^2 + p_\m p_\n)\,e^{ipx} \ ,
\label{ftaudef}
\end{eqnarray}
which implies (see Sec.~\ref{matchT} below)
\begin{equation}
  \bra{0} S_\m(x) \ket{\t} = ip_\m \hf_\t \,e^{ipx} \ .
\label{Smuftau}
\end{equation}
Here $\ket{\cdot}$ denotes a single-particle state of momentum $p$.
These states can be labeled by the effective fields of the
low-energy theory: $\p^a(x)$ is the pion field, which is related to the
non-linear field as
\begin{equation}
  \S = \exp(2i\p/\hf_\p) = \exp(2i\p^a T^a/\hf_\p) \ ,
\label{Spi}
\end{equation}
while $\t(x)$ is a new effective field for the dilatonic meson.
The indices $a,b=1,\ldots,N_f^2-1$ label the adjoint representation
of $SU(N_f)$.
Both the non-singlet axial currents $J_{5\m}^a$ and the energy-momentum
tensor $T_{\m\n}$ are renormalization-group invariant, and the same
applies to the decay constants $\hf_\p$ and $\hf_\t$.\footnote{%
  By contrast, the $U(1)_A$ current is not renormalization-group invariant
  because of the mixing of $\partial_\m J_{5\m}$ with $\tr F\tilde{F}$.
  This complicates the dependence of the effective theory on
  the renormalization scale of the microscopic theory \cite{KL}.
}

The large-$N$ scaling of the decay constants
follows from considering the two-point functions of $J_{5\m}^a$ or $T_{\m\n}$,
which gives
\begin{eqnarray}
  \hf_\p & \sim & \sqrt{N_c} \ ,
\label{fpiN}\\
  \hf_\t & \sim & N_c \ .
\label{ftauN}
\end{eqnarray}
Equation~(\ref{fpiN}) is recognized as
the familiar scaling law of the decay constant of any meson
made of fundamental-representation fermions.\footnote{%
  $\hf_\p$ is independent of $N_f$ because the flavor index
  in Eq.~(\ref{fpidef}) is held fixed.
}
With these scaling laws in mind we may thus set the dynamical infrared
scale $\L_{IR}$ to be
\begin{equation}
  \L_{IR} \sim \frac{4\p \hf_\p}{\sqrt{N_c}} \sim \frac{4\p \hf_\t}{N_c} \ .
\label{LIR}
\end{equation}
Here we use that $4\p \hf_\p$ is usually taken to be the ultraviolet cutoff
of the chiral lagrangian, it being the typical scale for the masses
of non-Nambu--Goldstone mesons.
The $\sim$ symbol in Eq.~(\ref{LIR})
serves to indicate that either expression can be used
as the definition of $\L_{IR}$.  The two options are equally good, because
in the Veneziano limit we expect the dimensionless ratio
$\sqrt{N_c}\, \hf_\p/\hf_\t$ to smoothly approach a non-zero value.
We recall that meson masses are $O(1)$
in the large-$N$ counting, and hence, with the exception of pNGBs,
we expect all meson masses to scale like $\L_{IR}$.

Turning to the light-meson sector, the three small parameters of the
low-energy expansion will be
\begin{subequations}
\label{pc}
\begin{eqnarray}
  && m/\L_{IR}\ ,
\label{pca}\\
  && 1/N \ ,
\label{pcb}\\
  && n_f - n_f^* \ .
\label{pcc}
\end{eqnarray}
\end{subequations}
As already mentioned, while $m/\L_{IR}$ and $1/N$ (equivalently, $1/N_c$)
can be taken arbitrarily small independent of each other,
the same is not true for $n_f - n_f^*$, which cannot be parametrically
smaller than $1/N$.  In the case where the expansion parameters
are comparable in size, we will again use $\d$ as a generic name
for any one of them.  For example, the behavior of the pion mass $m_\p$ is
\begin{equation}
  m_\p^2/\L_{IR}^2 \ \sim \
  (m/\L_{IR})^1 \, (1/N)^0 \, (n_f-n_f^*)^0 \ \sim \
  m/\L_{IR} \ \sim \ \d \ .
\label{mpiscale}
\end{equation}
In addition to the expansion in the input parameters of Eq.~(\ref{pc}),
the effective theory involves, as usual, a derivative expansion,
with $p_i \cdot p_j/\L_{IR}^2\sim \d$,
where $p_i \cdot p_j$ is the product of any two external momenta.

We next reexamine the anomalous divergence of the dilatation current,
which, in the chiral limit, reads
\begin{equation}
  \partial_\m S_\m  = - \tb\, [F^2]  \ ,
\label{dSm0}
\end{equation}
where, in terms of the 't Hooft coupling
\begin{equation}
  \a=\frac{g^2 N_c}{4\p} \ ,
\label{tHooft}
\end{equation}
we have (compare Eq.~(\ref{beta}))
\begin{equation}
  \tb % = \frac{\b(\a)}{4\a}
  = \frac{\m}{4\a} \frac{\partial \alpha}{\partial\mu} \ .
\label{betatilalph}
\end{equation}
Both $\a$ and $\tb(\a)$ are $O(1)$ in large-$N$ counting.

We will assume that the trace anomaly can be made
arbitrarily small by approaching the sill of the conformal window from below.
More specifically, we will assume that, in the chiral limit
(compare Eq.~(\ref{rentracec})),
\begin{equation}
  \tb[F^2] \sim (n_f-n_f^*)^\eta \ , \qquad n_f \nearrow n_f^* \ ,
\label{tbg}
\end{equation}
for some $\eta>0$, in the sense that all matrix elements of
the operator on the left-hand side vanish like $(n_f-n_f^*)^\eta$
when probed on the infrared scale $\L_{IR}$.
The main motivation for conjecturing Eq.~(\ref{tbg}) is the ever slower running
of the coupling as $N_f$ is increased towards the conformal window,
evidence for which comes from numerous lattice studies
(see Ref.~\cite{TD} and references therein).

Further heuristic support for Eq.~(\ref{tbg}) comes from modeling
chiral symmetry breaking by the gap equation.
According to this analysis, chiral symmetry breaking
in a walking theory sets in when the coupling has reached a critical value
$g_c^2=4\p^2/(3C_2)$ \cite{BH,YBM,AY,AKW}.
For the fundamental representation,
$C_2=(N_c^2-1)/(2N_c)$, and in the large-$N_c$ limit the
critical value of the 't Hooft coupling~(\ref{tHooft}) is $\a_c=2\p/3$.
In order to determine when the conformal window is entered,
$\a_c$ is compared with the (putative) value of the IRFP, $\a_*$
\cite{ATW,MY,DS}.\footnote{%
  For a different approach to this problem, see Ref.~\cite{BG}.
}
The latter is inferred, for example, from the two-loop beta function,
while ignoring the possible onset of spontaneous symmetry breaking
(see App.~\ref{VZlim}).
It follows that, if $\a_*>\a_c$, the IRFP is not reached,
and the theory exists in a
chirally broken (and confining) phase; whereas if $\a_*<\a_c$, the
running comes to a halt as $\a(\m)\nearrow\a_*$,
and the infrared theory is conformal.
In the Veneziano limit, $\a_*$ is a function of $n_f$,
and the conformal window is entered when $\a_*(n_f)=\a_c$.
Therefore, when the boundary separating the broken and conformal ``phases''
is reached from within the broken phase,
chiral symmetry breaking occurs at a coupling where the beta function
tends to zero.

We stress that we have no way of rigorously establishing the correctness
of Eq.~(\ref{tbg}).  This is the main weakness of our treatment.\footnote{%
  We would like to point out that the results of Ref.~\cite{MtsY} are in
  disagreement with our main assumption~(\ref{tbg}).  However, the analysis
  of Ref.~\cite{MtsY} involves a number of ingredients that do not constitute
  a systematic approximation, and, therefore, we believe that it does not
  disprove our assumption.
}
In addition,
unlike the parameters $m$ and $1/N_c$, in which we expand around zero,
we do not know the critical value $n_f^*$
around which we are to expand in $n_f$.
In practice, one would have to determine both $\eta$
and $n_f^*$ empirically.  One way to do that is by treating $\eta$ and $n_f^*$
as additional parameters of the low-energy effective theory,
which, together with the rest of the low-energy constants, are to be determined
by the usual procedure of comparing the correlation functions
of the microscopic theory with the predictions of the low-energy theory.

In order to avoid cumbersome notation, we will henceforth assume
that $\eta=1$ in Eq.~(\ref{tbg}).  Except for the trivial replacement of $n_f-n_f^*$
by $(n_f-n_f^*)^\eta$, nothing else would change in our analysis in
case that the actual value of $\eta$ is different.\footnote{%
  For the two-loop beta function, as well as for a crude model estimate
  of Eq.~(\ref{tbg}), see App.~\ref{VZlim}.
}

We conclude this subsection with several comments.

First, the reason why
we consider only the fundamental representation is that, for any other
representation,
asymptotic freedom will be lost if we take the large-$N_c$ limit
while increasing the number of higher-representation flavors
such that $N_f/N_c$ is held fixed.

Second, the one-loop mass anomalous dimension is $\g_m = 6 g^2 C_2/(16\p^2)$.
Since $C_2=(N_c^2-1)/(2N_c)$ for the fundamental representation,
the mass anomalous dimension is $O(1)$ in large-$N$ counting.
Unlike the beta function, there is no reason that the mass anomalous dimension
will become parametrically small near the bottom of the conformal window.
Indeed, numerical simulations of, \eg, the $N_f=8$ theory,
suggest that it is large \cite{CHPS} (see also Ref.~\cite{TD}).
A simplification that does take place is that,
since the running comes to a halt, $\g_m(\a(\m))$ is hardly changing over
many energy decades.  This has implications for the effective low-energy
theory, which we will discuss in Sec.~\ref{mutrans} below.

Finally, we recall that unlike in the usual large-$N_c$ limit,
in the Veneziano limit the axial anomaly does not vanish,
because we now have $g^2 N_f \sim 1$ in Eq.~(\ref{dJ5}).
For this reason, the flavor-singlet pseudoscalar meson does not become light,
and will not be considered in the rest of this paper.

%%%%%%%%%%%%%%%%%%%%%%%%%%%
%\newpage
\subsection{\label{match} Invariant potentials at tree level and the power counting}
%%%%%%%%%%%%%%%%%%%%%%%%%%%
The fields of the effective low-energy theory include the non-linear field
$\S(x)$ introduced in Sec.~\ref{Chisources}, and the dilatonic meson field
$\t(x)$.  The latter is inert under chiral transformations.
Imposing for the time being only the usual chiral power counting,
the leading-order
lagrangian consists of terms of order $p^2 \sim m \sim \d$,
and is given by\footnote{%
  We work in euclidean space.
}
\begin{equation}
  \tcl = \tcl_\p + \tcl_\t + \tcl_m + \tcl_d \ ,
\label{LeffL}
\end{equation}
where
\begin{eqnarray}
  \tcl_\p &=& \frac{f_\p^2}{4}\, V_\p(\t-\s)\, e^{2\t}
              \tr(\partial_\m \S^\dagger \partial_\m \S) \ ,
\label{LpV}\\
  \tcl_\t &=&
  \frac{f_\t^2}{2}\, V_\t(\t-\s)\, e^{2\t} (\partial_\m \t)^2  \ ,
\label{LtV}\\
  \tcl_m &=& -\frac{f_\p^2 B_\p}{2} \, V_M(\t-\s)\, e^{y\t}
  \tr\Big(\c^\dagger \S + \S^\dagger \c\Big) \ ,
\label{LmV}\\
  \tcl_d &=& f_\t^2 B_\t \, e^{4\t} V_d(\t-\s) \ .
\label{LdV}
\end{eqnarray}
$\tcl_\p$ and $\tcl_\t$ are the kinetic terms for pions
and for the dilatonic meson, respectively.  $\tcl_m$ is a generalized
chiral mass term, whereas $\tcl_d$ accounts for the self-interactions
of the dilatonic meson.  Actually, $\tcl_d$ is $O(\d^0)$, and not $O(\d^1)$,
in apparent violation of the rules of the effective theory.
We will see later on how this is resolved.
The relation of $f_\p$ and $f_\t$ to the physical decay constants
$\hf_\p$ and $\hf_\t$ in the chiral limit will be discussed below.
$B_\p$ and $B_\t$ are additional low-energy constants of mass dimension
one and two, respectively.  As we will see below (see Sec.~\ref{linVd}),
the dependence of the tree-level masses on the $B$ parameters
implies that they are $O(1)$ in large-$N$ counting.

The effective action $\tilde{S} = \int d^4x\, \tcl(x)$
is invariant under dilatations.
The transformation rules of the dynamical fields are
\begin{subequations}
\label{scleff}
\begin{eqnarray}
  \S(x) &\to& \S(\l x) \ ,
\label{scleffa}\\
  \t(x) &\to& \t(\l x) + \log \l \ .
\label{scleffb}
\end{eqnarray}
\end{subequations}
The transformation rule of $\s(x)$ is given in Eq.~(\ref{baredild}).
Notice the similarity of the transformation rules of
the dynamical field $\t(x)$ and the source field $\s(x)$.
The transformation rule of the chiral source $\c(x)$ that follows
from Eq.~(\ref{LmV}) is
\begin{equation}
  \c(x) \to \l^{4-y} \c(\l x) \ .
\label{transchiren}
\end{equation}
The transformation rule of the bare chiral source, given in Eq.~(\ref{baredilc}),
reflects its engineering dimension.  Compatibility of Eqs.~(\ref{transchiren})
and~(\ref{baredilc}) would require $y=3$.
In fact, the effective theory must be
constructed using the renormalized chiral source, because
it is the differentiation with respect to this renormalized
source that produces finite correlation functions of the microscopic theory.
The dilatation transformation rule of the renormalized chiral source
is different from that of the bare source.
We will derive it in Sec.~\ref{mutrans} below.

Because the shifts of $\s(x)$ and $\t(x)$ cancel each other,
the potentials $V_I$, $I=\p,\t,M,d,$ undergo only a coordinates
transformation
\begin{equation}
  V_I(\t(x)-\s(x))\to V_I(\t(\l x)-\s(\l x)) \ .
\label{transV}
\end{equation}
In other words, their scaling dimension is zero,
which is why we will sometimes refer to them as invariant potentials.
This transformation places no algebraic constraints on the
coefficients of the power-series expansion,
\begin{equation}
  V_I = \sum_{n=0}^\infty \frac{c_{I,n}}{n!}\, (\t-\s)^n \ .
\label{expandVI}
\end{equation}
The coefficients $c_{I,n}$ therefore amount to infinitely many,
linearly independent low-energy parameters.

In this subsection, we will establish a power-counting hierarchy
\begin{equation}
  c_{I,n} \sim (n_f - n_f^*)^n \sim \d^n \ .
\label{cIn}
\end{equation}
This will allow us to set the potentials $V_\p$, $V_\t$ and $V_M$
equal to one in the
leading-order lagrangian, because they multiply terms
that are already $O(\d)$;
and to truncate $V_d$ to a linear function, as in Eq.~(\ref{Vd01}).

When writing down the chiral lagrangian, it is customary to keep explicit
the $m$ dependence.  Similarly, the $n_f-n_f^*$ dependence can be made
explicit by performing the further expansion\footnote{%
  Recall that we are assuming $\eta=1$ in Eq.~(\ref{tbg}), see Sec.~\ref{conj}.
}
\begin{equation}
  c_{I,n} = \sum_{k=0}^\infty \tc_{I,nk} (n_f-n_f^*)^k \ .
\label{expandcn}
\end{equation}
The new coefficients $\tc_{I,nk}$ are independent of all three
expansion parameters $m$, $1/N$ and $n_f-n_f^*$.  The power counting~(\ref{cIn})
can be equivalently stated as
\begin{equation}
  \tc_{I,nk} = 0 \ , \qquad k<n \ .
\label{Cnk}
\end{equation}
Below, we will mostly refer to the $c_{I,n}$ coefficients,
but we will keep in mind
that each of them is a function of $n_f-n_f^*$.

The hierarchy~(\ref{cIn}) is analogous to the one established
in Refs.~\cite{GL2,KL} for the coefficients of the invariant potentials
$V(\eta-\th)$ discussed in Sec.~\ref{Chisources},
which occur in the low-energy effective theory
that accommodates the $U(1)_A$ symmetry.  The main difference is
that the power-series coefficients of $V(\eta-\th)$ are suppressed
by powers of $1/N_c$ in the usual large-$N_c$ limit.
By contrast, here the power-series coefficients of
the $V_I(\t-\s)$ potentials are suppressed by powers of $n_f-n_f^*$,
which is a new small parameter that is present in the Veneziano limit
but not in the usual large-$N_c$ limit.

We will derive Eq.~(\ref{cIn}) by matching judiciously chosen
correlation functions of the low-energy theory to the corresponding
correlation functions of the microscopic theory, in the chiral limit.
In order to understand the physical origin of Eq.~(\ref{cIn}),
we begin by considering correlations functions of the microscopic theory
obtained by differentiating $n$ times with respect to the $\s(x)$ source field.
As usual, the role of the external sources is to generate operator insertions,
and thus in the microscopic theory we will assume that no two differentiations
are done at coinciding points.  We thus have
\begin{subequations}
\label{Gn}
\begin{eqnarray}
  \GMIC_n(x_1,\ldots,x_n) &=& (-1)^n
  \frac{\partial}{\partial \s(x_1)}\cdots\frac{\partial}{\partial \s(x_n)}\, W
\label{Gna}\\
  &=& \svev{T_{an}(x_1)\cdots T_{an}(x_n)}_{m=0}
\label{Gnb}\\
  &=& \tb^n \svev{F^2(x_1)\cdots F^2(x_n)} \ .
\label{Gnc}
\end{eqnarray}
\end{subequations}
Here $W=\log Z$.  It is understood that all source fields
are set to zero after the differentiations are done (since we work
in the chiral limit, this includes setting $\cs(x)=\cp(x)=0$,
{\it cf.} Eq.~(\ref{spurvev})).
Allowing also for differentiations with respect to the scalar and pseudoscalar
sources $\cs_{AB}$ and $\cp_{AB}$ gives us access to
\begin{eqnarray}
  && \hspace{-8ex}
  \GMIC_{nk\ell}(x_1,\ldots,x_n;y_1,\ldots,x_k;z_1,\ldots,z_\ell) \ =
\label{GnSP}\\
  &=& (-1)^{n+k+\ell}
  \frac{\partial}{\partial \s(x_1)}\cdots\frac{\partial}{\partial \s(x_n)} \;
  \frac{\partial}{\partial \cs(y_1)}\cdots\frac{\partial}{\partial \cs(y_k)} \;
  \frac{\partial}{\partial \cp(z_1)}\cdots\frac{\partial}{\partial \cp(z_\ell)}
  \, W
\NON
  &=& \tb^n \svev{F^2(x_1)\cdots F^2(x_n)\;\bj\j(y_1)\cdots \bj\j(y_k)\;
  \bj i\g_5\j(z_1)\cdots \bj i\g_5\j(z_\ell)} \ ,
\nonumber
\end{eqnarray}
where we have suppressed the flavor indices.  We see that, by Eq.~(\ref{tbg}),
\begin{equation}
  \GMIC_{nk\ell} =  O(\d^n) \ .
\label{Gnpc}
\end{equation}
Here and in the rest of this subsection, $O(\d^n)$
will always mean $O((n_f-n_f^*)^n)$.
In Eq.~(\ref{GnSP}), all the operators are renormalized ones.\footnote{%
  To avoid cumbersome notation, we have omitted the square brackets
  used in Sec.~\ref{mic} to denote renormalized operators.
}
Correspondingly, the chiral sources used to generate $\GMIC_{nk\ell}$
are renormalized, too.

Turning to the low-energy effective theory, in order to probe the
potentials $V_I(\t-\s)$, we will carry out multiple differentiations
at the {\it same} spacetime point.  This immediately raises the issue
of reconciling this procedure with the avoidance of the limit
of coinciding points in the correlation functions of the microscopic theory.
There is in fact no conflict.
The point is that coinciding spacetime points do not have
the same physical significance in the microscopic and the effective
theories.  While the pNGBs are represented in the effective theory
as point particles, in reality they have a finite size,
roughly of order $1/\L_{IR}$.
We will thus choose to consider operator insertions
in the microscopic theory at non-coinciding points,
but at distances much smaller than $1/\L_{IR}$, such that they
collapse to insertions at coinciding points in the effective low-energy theory.
In addition, asymptotic states are obtained from operator insertions at
large distances, in the same way in the microscopic and the effective theories.
This will allow us
to enforce the power counting of Eq.~(\ref{Gnpc}) on the effective theory's
correlation functions that we will need in order to probe the
$V_I(\t-\s)$ potentials.

Among the correlation functions obtained in the effective theory
by differentiating multiple times with respect to $\s(x)$ at the same
spacetime point, there is a special subset, in which all the differentiations
are applied to the same potential $V_I$.  This gives rise to a diagram
containing the $n$-th derivative of $V_I$, which,
when evaluated at $\s=0$, is simply (compare Eq.~(\ref{expandVI}))
\begin{equation}
  \bigg(\!-\frac{\partial}{\partial\s}\bigg)^n\; V_I \bigg|_{\s=0}
  = \sum_{k=0}^\infty \frac{c_{I,n+k}}{k!}\, \t^k = V_I^{(n)}(\t)\ .
\label{dVn}
\end{equation}
The natural way for the correlation functions of the effective theory
to satisfy Eq.~(\ref{tbg}), is that the power-series coefficients of $V_I$
satisfy
\begin{equation}
  c_{I,k} = O(\d^n) \ , \qquad k \ge n \ ,
\label{cIkn}
\end{equation}
because this implies that $V_I^{(n)}(\t)= O(\d^n)$ in its entirety.
It is easy to see that conditions~(\ref{cIn}),~(\ref{Cnk})
and~(\ref{cIkn}) are mathematically equivalent.  As it turns out,
it is convenient to establish the power-counting hierarchy
in the form of Eq.~(\ref{cIkn}).

There are a few more technical points we need to address before turning
to the actual derivation.  First, the zeroth-order term in the double expansion
of each potential, $\tc_{I,00}$, is redundant, because each term in
the lagrangian~(\ref{LeffL}) is already multiplied by a low-energy constant.
We will use this freedom to set $\tc_{M,00}=\tc_{\p,00}=\tc_{\t,00}=1$.
The role of $\tc_{d,00}$, the leading-order term in the expansion of $V_d$,
is more subtle.  The $\tc_{d,00}$-dependent term
is the only part in the entire lagrangian which is $O(1)$.
All other terms start at $O(\d)$, which is how the leading-order lagrangian
normally scales according to the low-energy power counting.
As we will show in Sec.~\ref{linVd} below,
when we develop the diagrammatic expansion
for each theory, there is a redefinition of the $\t$ field that
allows us to set $c_{d,0}=-c_{d,1}/4$.  This is equivalent
to setting $\tc_{00}=0$ and $\tc_{01}=-\tc_{11}/4$, so that,
after the redefinition, the entire leading-order lagrangian is $O(\d)$.

For the diagrammatic expansion, we split the dilatonic meson field as
\begin{equation}
  \t(x) = v +\tilt(x)/(e^v f_\t) \equiv v +\tilt(x)/\hf_\t \ ,
\label{cnnprop}
\end{equation}
where $v=\svev{\t}$ is the classical solution, and
the quantum field $\tilt(x)$ has a canonically normalized kinetic term.
Applying the $\t$ field redefinition of Sec.~\ref{linVd} to Eq.~(\ref{cnnprop})
shifts the classical solution to $v=0$ in the chiral limit,
and the dependence on the original value of $v$ gets absorbed
into the decay constants,
\begin{subequations}
\label{fphys}
\begin{eqnarray}
  \hf_\p &=& e^v f_\p \ ,
\label{fphysa}\\
  \hf_\t &=& e^v f_\t \ ,
\label{fphysb}
\end{eqnarray}
\end{subequations}
and into the two other low-energy constants,
\begin{subequations}
\label{Bphys}
\begin{eqnarray}
  \hB_\p &=& e^{(y-2)v} B_\p \ ,
\label{Bphysa}\\
  \hB_\t &=& e^{2v} B_\t \ .
\label{Bphysb}
\end{eqnarray}
\end{subequations}
In these equations, $v$ is the classical solution in the chiral limit before
the field redefinition,
given by Eq.~(\ref{minV}) below.  After the $\t$ field redefinition,
which sets $v=0$ in the chiral limit,
it can moreover be shown that the ground state is stable against
the inclusion of higher-order terms in the classical potential.\footnote{%
  In fact, for what we need here it would be enough to set
  $\tc_{d,00}=0$, leaving $\tc_{d,01}$ free; see Sec.~\ref{lagnlo}.
}

As for the non-linear $\S$ field,
in this paper we assume that $m>0$ (or, if the chiral limit has been taken,
that it was approached as $m\searrow 0$), hence its classical vacuum
expectation value is $\svev{\S}=1$.
The diagrammatic expansion is then generated as usual by expanding $\S$
in terms of the pion field, Eq.~(\ref{Spi}).

\medskip

We will prove the power-counting hierarchy~(\ref{cIkn})
using the diagrammatic expansion described above.
The proof of Eq.~(\ref{cIkn}) proceeds by induction.
We assume that Eq.~(\ref{cIkn}) has been proved up to $n-1$, so that
$c_{I,k}=O(\d^k)$ for all $k\le n-1$, and $c_{I,k}=O(\d^{n-1})$ for $k\ge n-1$.
In the induction step we must therefore improve the bound
to $c_{I,k}=O(\d^n)$ for $k\ge n$.

In the effective theory, the $n$ differentiations with respect to
the $\s(x)$ source will all be done at the same point.  We will be interested
in the special subset of diagrams mentioned above,
in which all $n$ differentiations have been applied to a single $V_I$,
so that the diagram involves a single vertex
\begin{equation}
  V_I^{(n)}(\t(x)) \ .
\label{Vn}
\end{equation}
We claim that the sum of diagrams with a single such vertex
must be $O(\d^n)$ by itself.  When we apply the $\s$ differentiations,
we will also obtain diagrams with the factorizable form
\begin{equation}
  V_{I_1}^{(k_1)}(\t(x)) \cdots V_{I_\ell}^{(k_\ell)}(\t(x)) \ ,
  \qquad k_1 + \cdots + k_\ell = n \ ,
\label{Vnosplit}
\end{equation}
where at least two $k_i$'s are larger than zero,
and therefore all $k_i<n$.
Now, the factorizable-form diagrams may be obtained by first using $\ell$
non-coinciding points to obtain
\begin{equation}
  V_{I_1}^{(k_1)}(\t(x_1)) \cdots V_{I_\ell}^{(k_\ell)}(\t(x_\ell)) \ ,
\label{Vsplit}
\end{equation}
and then taking the limit where all the $\ell$ spacetime points are
brought together.  Since $k_i<n$, the induction hypothesis implies
$V_{I_i}^{(k_i)}(\t(x_i))=O(\d^{k_i})$ for all $i$.
The diagrams with the point-split
structure~(\ref{Vsplit}) are thus $O(\d^n)$, and
the same is true in the limit of coinciding points.
Therefore, the left-over part, which is the (sum of) diagrams with the
non-factorizable vertex~(\ref{Vn}), must be $O(\d^n)$ by itself.

\medskip

The rest of the argument is more technical, and requires a case-by-case study.
We will give here the proof of Eq.~(\ref{cIkn}) for the power-series coefficients
of $V_d$, which is the potential that is easiest to handle.  The demonstration
of Eq.~(\ref{cIkn}) for the other three potentials $V_M$, $V_\p$ and $V_\t$
makes use of the same ingredients, but is slightly more involved,
and is relegated to App.~\ref{inductiveV}.
The reader who is not interested in these details may skip to the next
subsection.

Let us carry out no additional differentiations except for
the $n$ derivatives with respect to the $\s(x)$ field, and
isolate the diagrams with a non-factorizable vertex.
After dividing by $\hf_\t^2 \hB_\t$
we obtain the effective theory's correlation function
\begin{equation}
  \GEFT_{d,n} = (\hf_\t^2 \hB_\t)^{-1} \svev{\O_{d,n}+\O_{\p,n}+\O_{\t,n}} \ ,
\label{GeftVd}
\end{equation}
where
\begin{eqnarray}
  \O_{d,n} &=& \hf_\t^2 \hB_\t\, V_d^{(n)}(\t)\, e^{4\t} \ ,
\label{nvertexd}\\
  \O_{\p,n} &=& \frac{\hf_\p^2}{4}\, V_\p^{(n)}(\t)\,
  e^{2\t} \tr(\partial_\m \S^\dagger \partial_\m \S) \ ,
\label{nvertexp}\\
  \O_{\t,n} &=& \frac{\hf_\t^2}{2}\, V_\t^{(n)}(\t)\,e^{2\t} (\partial_\m \t)^2  \ .
\label{nvertet}
\end{eqnarray}
The generalized mass term $\tcl_m$ does not contribute, because we did not
differentiate with respect to the chiral source $\c(x)$,
but we did set $\c(x)=m=0$.  The division by $\hf_\t^2 \hB_\t$
in Eq.~(\ref{GeftVd})
makes $\GEFT_{d,n}$ dimensionless, and, in addition,
it removes the leading large-$N$ dependence, so that
$\GEFT_{d,n}$ is $O(1)$ in large-$N$ counting.

Our first step is to establish that $c_{d,n}=O(\d^n)$.
Considering the expansion
\begin{equation}
 \O_{d,n} =
 \sum_{k=0}^\infty \frac{c_{d,n+k}}{k!}\, \hf_\t^2 \hB_\t\, \t^k e^{4\t} \ ,
\label{expandOd}
\end{equation}
we need to find a suitable limit where
$\GEFT_{d,n}$ is dominated by the first term on the right-hand side
of Eq.~(\ref{expandOd}).  If we can do so, we will have
\begin{equation}
  \GEFT_{d,n} = c_{d,n} + O(\d^{n+1})\ ,
\label{GeftVdfinal}
\end{equation}
and the desired result will follow from the matching to the microscopic
theory.

We will derive Eq.~(\ref{GeftVdfinal}) by working in the parameter region
\begin{subequations}
\label{epsdelta}\\
\begin{eqnarray}
  m/\L_{IR} \ \sim \ 1/N \ \sim \ p_i\cdot p_j &\sim& \e_n \ ,
\label{epsdeltaa}\\
  n_f-n_f^* &\sim& \d \ ,
\label{epsdeltab}
\end{eqnarray}
where $p_i\cdot p_j$ denotes the product of any two external momenta,
and where
\begin{equation}
  \e_n \ll \d^n \ .
\label{epsdeltac}
\end{equation}
\end{subequations}
We begin with the observation that $\svev{\O_{\p,n}}$ contains a pion loop,
which, including the normalization factor from Eq.~(\ref{GeftVd}), is
\begin{equation}
  \frac{\hf_\p^2 \svev{\tr(\partial_\m \S^\dagger \partial_\m \S)}}
       {\hf_\t^2 \hB_\t}
  \sim
  \frac{\svev{\tr(\partial_\m \p \partial_\m \p)}}
       {\hf_\t^2 \hB_\t}
  \sim \frac{N^2 m_\pi^4}{\hf_\t^2 \hB_\t}
  \sim \frac{m_\pi^4}{\L_{IR}^4}
  \sim \e_n^2 \ ,
\label{piloop}
\end{equation}
where we have used that the number of pions scales like $N_f^2 \sim N^2$,
the large-$N$ scaling of $\hf_\t$ and $\hB_\t$, and Eq.~(\ref{Spi}).
It follows that $\svev{\O_{\p,n}}$ can be neglected.
We next consider $\svev{\O_{\t,n}}$.
Since we have shifted the classical vacuum to $v=0$,
we may substitute $\t=\tilt/\hf_\t$.  The contribution of $\svev{\O_{\t,n}}$
thus involves the dilatonic meson tadpole\footnote{%
  Note that $\svev{\tilt\partial_\m \tilt}=0$.
}
\begin{equation}
  \frac{\svev{\hf_\t^2 (\partial_\m \t)^2}}{\hf_\t^2 \hB_\t}
  = \frac{\svev{(\partial_\m \tilt)^2}}{\hf_\t^2 \hB_\t}
  \sim \frac{m_\t^4}{\hf_\t^2 \hB_\t} \sim \frac{m_\t^4}{N^2 \L_{IR}^4} \ .
\label{dtausq}
\end{equation}
As we will see in Sec.~\ref{linVd} below, $m_\t^2 \sim \L_{IR}^2 |n_f-n_f^*|$
in the chiral limit, and so, generically,
the suppression of dilatonic meson loop comes from the smallness of
both $|n_f-n_f^*|$ and $1/N$.
Here we invoke the parameter region~(\ref{epsdelta}),
because we must maintain the freedom to vary $n_f-n_f^*$.
Therefore, we cannot neglect $\svev{\O_{\t,n}}$
on the grounds of its $m_\t$ dependence.
However, Eq.~(\ref{dtausq}) also involves a factor of $1/N^2 \sim \e_n^2$,
and this does justify the neglect of $\svev{\O_{\t,n}}$.

At this stage we are left with $\svev{\O_{d,n}}$.  The last step
is to observe that, besides $c_{d,n}$, all other contributions
to $\svev{\O_{d,n}}$
involve some number of dilatonic meson tadpoles $\svev{\t^2}$,
each of which provides an additional suppression factor of $1/N^2$.
This completes the proof that $c_{d,n}=O(\d^n)$.

\medskip

In the above argument, loop diagrams were always subleading.
This is an example of a general result,
first derived by Weinberg in the context of the chiral lagrangian \cite{SW},
that the loop expansion of every correlation function in the effective theory
obeys the standard power counting.  Thus, every loop diagram is always
suppressed relative to the tree diagrams that contribute to a given
correlation function.  Weinberg's result was generalized to the case
where a dilatonic meson is present in Ref.~\cite{CT}.
While our setup is different from Ref.~\cite{CT},
both in terms of the expansion parameters~(\ref{pc}) and
the particular parameter range~(\ref{epsdelta}),
the argument remains essentially the same.
Henceforth, it will thus suffice to consider tree diagrams only.

In order to extend Eq.~(\ref{cIkn}) to $c_{d,k}$ for $k>n$, we will need
$\ell=k-n$ external $\t$ legs.  These external legs must be generated
by differentiating with respect to one of the source fields, since
Eq.~(\ref{GnSP}) gives the complete set of correlation functions of the
microscopic theory that can be matched to the effective theory.
We will differentiate with respect to (the flavor trace of)
the scalar source $\cs(y)$.
Including a suitable normalization factor and a trivial subtraction, we let
\begin{subequations}
\label{Ocs}
\begin{eqnarray}
  \O_\cs &=& -\frac{1}{\hf_\p^2 \hB_\p} \frac{\partial \tcl}{\partial \cs}
\label{Ocsa}\\
  &=& \half \, V_M(\t)\, e^{y\t} \tr (\S+\S^\dagger) \ ,
\NON
  \tO_\cs &=& \O_\cs - N_f
\label{Ocsb}\\
  &=& \frac{y\tilt}{\hf_\t/N_f} - \frac{2}{\hf_\p^2}\tr(\p^2) + \cdots \ ,
\nonumber
\end{eqnarray}
\end{subequations}
where we have used Eq.~(\ref{LmV}) and $\tc_{M,00}=1$.  This shows that $\tO_\cs$
can be used as an interpolating field for the dilatonic meson.
As a generalization of Eq.~(\ref{GeftVd}), we will consider
$\tGEFT_{d,n,\ell}(p_1,\ldots,p_\ell)$, the Fourier transform of
\begin{equation}
  \GEFT_{d,n,\ell}
  = (\hf_\t^2 \hB_\t)^{-1} \svev{(\O_{d,n}(x)+\O_{\p,n}(x)+\O_{\t,n}(x)) \,
  \tO_\cs(y_1)\cdots\tO_\cs(y_\ell)} \ .
\label{GeftVdk}
\end{equation}
The demonstration that $c_{d,k}=O(\d^n)$ for $k>n$ will proceed by an
``inner'' induction.  Before turning to the general case,
let us examine the two simplest cases.

We start with $\ell=k-n=1$.  The leading contribution to $\GEFT_{d,n,1}$
is a tree diagram with a single dilatonic meson line connecting
$\O_{d,n}(x)$ with $\tO_\cs(y_1)$.
Expanding $V_d^{(n)}(\t)\, e^{4\t} = c_{d,n} + \t(4c_{d,n}+c_{d,n+1}) + O(\t^2)$
we find that, after amputation, the contribution to $\tGEFT_{d,n,1}$ is
equal to $4c_{d,n}+c_{d,n+1}$ up to an $O(1)$ factor.
Since we already know that $c_{d,n}=O(\d^n)$,
the conclusion is that we must have $c_{d,n+1}=O(\d^n)$ as well.

Proceeding to $\ell=k-n=2$, now $\svev{\O_{d,n}(x)\,\tO_\cs(y_1)\tO_\cs(y_2)}$
receives contributions from three topologically distinct tree diagrams,
each containing two dilatonic meson lines.
The tree diagram we need is the one in which $\O_{d,n}(x)$
is connected by a line to each of $\tO_\cs(y_1)$ and $\tO_\cs(y_2)$.
In addition, there two more tree-diagram topologies. In one case,
the two lines emanate from $\tO_\cs(y_1)$, and connect it to $\O_{d,n}(x)$
and to $\tO_\cs(y_2)$.  In the other case, the roles
of $\tO_\cs(y_1)$ and $\tO_\cs(y_2)$ are interchanged.
At this point we use the external momenta as probes.
Each tree-diagram topology has a unique dependence on the external momenta,
and this allows us to isolate the tree diagram that we are after,
the one in which the two lines emanate from $\O_{d,n}(x)$.
After amputation, the result involves a linear combination
of $c_{d,n}$, $c_{d,n+1}$ and $c_{d,n+2}$, and, since we already know
that $c_{d,n}$ and $c_{d,n+1}$ are $O(\d^n)$, it follows that
the same is true for $c_{d,n+2}$.

The structure of the induction step is now clear.
Having established that $c_{d,k}=O(\d^n)$ for $n\le k \le n+\ell-1$,
we will in the next step consider $\tGEFT_{d,n,\ell}(p_1,\ldots,p_\ell)$.
By Weinberg's theorem (and its generalizations), tree diagrams
make the leading contribution.  It is easily seen that $\O_{\p,n}$
cannot contribute to any tree diagram, and can thus be neglected.\footnote{%
  Note that $\tO_\cs$ produces at least two pions.
}
The case of $\O_{\t,n}$ is different.  For $\ell\ge 2$,
there are tree diagrams that receive contributions from both $\O_{d,n}$
and $\O_{\t,n}$.  However, due to the presence of two derivatives,
in all cases the contribution of $\O_{\t,n}$ will be suppressed
by a relative factor $p_i\cdot p_j \sim \e_n$.  We are left with
the tree diagrams constructed only from $\O_{d,n}$.  Once again
using the different momentum dependence we isolate the tree diagram
in which the $\ell$ lines all emanate from $\O_{d,n}$, and, finally,
using the induction hypothesis we conclude that $c_{d,n+\ell}=O(\d^n)$.

For the generalization of Eq.~(\ref{cIkn}) to the other leading-order potentials
$V_M$, $V_\p$ and $V_\t$, as well as for a brief discussion of the potentials
encountered at the next-to-leading order and beyond, the reader is referred
to App.~\ref{inductiveV}.

%%%%%%%%%%%%%%%%%%%%%%%%%%%
%\newpage
\subsection{\label{mutrans} The leading-order lagrangian}
%%%%%%%%%%%%%%%%%%%%%%%%%%%
Having established the power counting for the coefficients of the
invariant potentials, we are done with the $\s(x)$ source field,
and we set it equal to zero.
The power counting~(\ref{cIkn}) allows us to replace $V_M$, $V_\p$ and $V_\t$
by one in the leading-order lagrangian,
and to truncate $V_d$ to a linear function, as in Eq.~(\ref{Vd01}).
The result is the following leading-order effective lagrangian,
\begin{equation}
  \cl = \cl_\p + \cl_\t + \cl_m + \cl_d \ ,
\label{Leff}
\end{equation}
where\footnote{%
  In the rest of this paper we reserve the abbreviations $c_n$ and $\tc_{nk}$
  for $c_{d,n}$ and $\tc_{d,nk}$, respectively.
}
\begin{eqnarray}
  \cl_\p &=& \frac{f_\p^2}{4}\, e^{2\t}
            \tr(\partial_\m \S^\dagger \partial_\m \S) \ ,
\label{Lp}\\
  \cl_\t &=& \frac{f_\t^2}{2}\, e^{2\t} (\partial_\m \t)^2  \ ,
\label{Lt}\\
  \cl_m &=& -\frac{f_\p^2 B_\p}{2} \, e^{y\t}
  \tr\Big(\c^\dagger \S + \S^\dagger \c\Big) \ ,
\label{Lm}\\
  \cl_d &=& f_\t^2 B_\t \, e^{4\t} (c_0 + c_1\t) \ .
\label{Ld}
\end{eqnarray}
In spite of the absence of the $\s$ field,
the effective action $S = \int d^4x\, \cl(x)$
is invariant under a different scale transformation,
where the dynamical effective fields transform in the same way as before,
and, in addition,
\begin{subequations}
\label{glbltrans}
\begin{eqnarray}
  c_0 &\to& c_0 - c_1 \log\l \ ,
\label{glbltransa}\\
  c_1 &\to& c_1 \ ,
\label{glbltransb}
\end{eqnarray}
\end{subequations}
where we have promoted the coefficients of $V_d(\t)$ to global spurions.
[As shown in App.~\ref{diltc}, this invariance can be generalized
to the case that we set $\s(x)=0$ in the lagrangian~(\ref{LeffL})
while keeping the potentials $V_I(\t)$ completely arbitrary.]

There is one more issue to address.  As already mentioned,
the effective theory must be
constructed using the renormalized chiral source.
We will now derive the dilatation, or scale, transformation
of the renormalized chiral source field $\c$.  As we will see, this
fixes the value of $y$ in Eq.~(\ref{transchiren}) in terms of the
mass anomalous dimension.

We start by reexamining the bare microscopic partition function in the presence
of the $\s$ field.  This partition function is in fact
invariant under a family of scale transformations
that depend on a continuous parameter $\z$.  The transformation rules
of the dynamical and source bare fields, Eqs.~(\ref{baredila})
though~(\ref{baredilc}), remain the same as before,
whereas Eqs.~(\ref{baredild}) and~(\ref{baredile}) are generalized to
\begin{subequations}
\label{splitmusig}
\begin{eqnarray}
  \s(x) &\to& \s(\l x) + \z \log \l \ ,
\label{splitmusiga}\\
  \m &\to& \l^{(1-\z)} \m  \ ,
\label{splitmusigb}\\
  \tg_0 &\to& \tg_0\ .
\label{splitmusigc}
\end{eqnarray}
\end{subequations}
The new, dimensionless, parameter $\tg_0$ is defined in terms of
the bare coupling $g_0$ and the renormalization scale $\m$,
according to
\begin{equation}
  \tg_0 = \m^{d/2-2} g_0 \ .
\label{g0}
\end{equation}
The invariance of the microscopic partition function under this family
of scale transformations follows from the fact that
\begin{equation}
  \m e^{\s(x)} \to \l \m e^{\s(\l x)} \ ,
\label{musigma}
\end{equation}
independent of $\z$.  The transformation~(\ref{baredil}) is recovered
for the special case $\z=1$.
Once we set $\s(x)=0$, the only consistent choice in Eq.~(\ref{splitmusig})
is $\z=0$.  The entire scale transformation~(\ref{musigma})
is then ``carried'' by the renormalization scale $\m$,
which transforms according to its engineering dimension, $\m \to \l \m$.
This is the first ingredient we need.

Our $\z=0$ scale transformation has something in common
with a renormalization group transformation,
because the renormalization scale $\m$ changes under it in the same way.
What distinguishes this scale transformation from a
renormalization-group transformation, is that,
in a renormalization-group transformation, the bare fields and parameters
are kept invariant; whereas under our scale transformation
all bare fields and parameters transform according to their engineering
dimension.

In particular, the bare coupling $g_0$ transforms non-trivially
under our scale transformation, in accordance with the fact
that it is dimensionful for $d\ne 4$.
The dimensionless $\tg_0$ is invariant.
Because $Z(g)g=  \m^{d/2-2} g_0= \tg_0$, it follows that
the renormalized coupling $g$, too, is held fixed.
Postponing momentarily which value of the renormalized coupling we
should choose, the very fact
that it is kept invariant allows us to obtain the transformation rule
of the renormalized fermion mass $m$ and chiral source $\c(x)$, which is
\begin{equation}
  \c(x) \to \l^{1+\g_m} \c(\l x) \ .
\label{transchi}
\end{equation}
We see that the scale transformation now involves
both the engineering and the anomalous dimension.  The mass anomalous dimension
is to be evaluated at the fixed value of the coupling, $\g_m=\g_m(g)$,
and we thus arrive at the simple power in Eq.~(\ref{transchi}) above.
Requiring compatibility of this new transformation rule with Eq.~(\ref{transchiren})
we conclude that
\begin{equation}
  y = 3 - \g_m \ .
\label{gammay}
\end{equation}

The final remaining question is then what
value of $\g_m$, or, equivalently, $g$, we should choose
to fix $y$ in the effective theory.
By construction, the effective low-energy theory is an expansion
in $m$ and $n_f-n_f^*$.\footnote{%
  For this discussion, the expansion in $1/N$ only
  plays the role of allowing $n_f-n_f^*$ to tend to zero continuously.
}
The starting point of the expansion is therefore $n_f=n_f^*$ (and $m=0$).
Let $\a_*^c$ be the location of the infrared fixed point, expressed
in terms of the 't Hooft coupling~(\ref{tHooft}),
as the conformal window is entered.
Assuming that the beta and gamma functions depend
continuously on $n_f$, at $n_f=n_f^*$ the right choice for the argument of
$\g_m$ is $\a=\a^c_*$, and thus,
\begin{equation}
  \g_m = \g_m(\a_*^c) \equiv \g_m^* \ .
\label{gammastar}
\end{equation}
The physical picture behind this choice is that, when $n_f$ tends to $n_f^*$
from below, chiral symmetry breaking will set in at a scale where,
first, the beta function has become vanishingly small, and second,
the value of the renormalized coupling tends to $\a_*^c$.
Corrections to Eq.~(\ref{gammastar}) can be expanded
in $n_f-n_f^*$, and thus they are subsumed in the $V_I$ potentials
that fully account for the dependence of the low-energy theory on $n_f-n_f^*$.

Numerical simulations find that the mass anomalous dimension is,
within error, always in the range (see Ref.~\cite{TD} and references therein)
\begin{equation}
  0\le \g_m \leqx 1 \ ,
\label{grange}
\end{equation}
suggesting that the actual value of $y$ should be somewhere in the range
\begin{equation}
  2\leqx y \le 3 \ .
\label{yrange}
\end{equation}
For theoretical work supporting a value of $\g_m$
near the upper bound in Eq.~(\ref{grange})
for walking theories, see Refs.~\cite{YBM,AY,AKW,CG,LE}.

%%%%%%%%%%%%%%%%%%%%%%%%%%%
%\newpage
\section{\label{LO} The effective theory at tree level}
%%%%%%%%%%%%%%%%%%%%%%%%%%%
In this section we study the classical vacuum and the tree-level spectrum.
When writing down the leading-order lagrangian~(\ref{Leff}) we have already
set $\s(x)=0$.  The last preparatory step is to set the chiral source
to its ``expectation value'' $\c(x)=m$.
We will assume that either $m>0$, or else the chiral limit has been taken
as $m\searrow 0$, so that, in all cases, the classical pion vacuum is $\S=1$
or, equivalently, $\tr\svev{\S}=N_f$.

The vacuum of the dilatonic meson is determined by minimizing
the classical potential
\begin{subequations}
\label{Vtaueff}
\begin{equation}
  V_{cl}(\t) = f_\t^2 B_\t U(\t) \ ,
\label{Vtaueffa}
\end{equation}
where
\begin{eqnarray}
  U(\t) &=& -\frac{f_\p^2 B_\p N_f m}{f_\t^2 B_\t}\, e^{y\t} + V_d(\t) e^{4\t}
\label{Vtaueffb}\\
  &=&  -\frac{m}{\cm}\, e^{y\t} + V_d(\t) e^{4\t} \ ,
\NON
  \cm &\equiv& \frac{f_\t^2 B_\t}{f_\p^2 B_\p N_f} \ .
\label{Vtaueffc}
\end{eqnarray}
\end{subequations}
Here $\cm$ has the dimensions of mass.  The rescaled potential
$U(\t)$ is dimensionless, and is  $O(1)$ in large-$N$ counting.
This follows from the scaling laws of the decay constants
(see Sec.~\ref{conj}) and the fact that $B_\p$ and $B_\t$ are themselves
$O(1)$ in large-$N$ counting.  The latter property follows from the
dependence of the pion and dilatonic meson masses on
these parameters (see Sec.~\ref{linVd} below), taking into account that
meson masses are $O(1)$ in large-$N$ counting.
The pion mass is parametrically small in $m$,
and dilatonic meson mass is parametrically small in $n_f-n_f^*$ and $m$
combined, while neither mass is parametrically small in $1/N$.\footnote{%
  As discussed in Sec.~\ref{conj}, the Veneziano large-$N$ limit is needed
  in order to enable us to treat $n_f-n_f^*$ as a continuous parameter
  that can be made arbitrarily small.
}

The function $V_d(\t)$ is linear.  If we make use of Eq.~(\ref{expandcn})
and the power-counting hierarchy~(\ref{Cnk}) to keep only the
$O(n_f-n_f^*)$ term in its expansion, we get
\begin{equation}
  V_d(\t) = c_0 + c_1 \t
  = \tc_{00} +  (n_f-n_f^*)(\tc_{01} + \tc_{11} \t) \ ,
\label{Vdnf}
\end{equation}
where the low-energy constants $\tc_{00}$,  $\tc_{01}$, and $\tc_{11}$
are independent of all three expansion parameters in Eq.~(\ref{pc}).

This section is organized as follows.  In Sec.~\ref{constVd} we study
the classical ground state approximating $V_d$ by its leading constant term.
We find that this truncation does not give rise to a consistent starting point
for the diagrammatic expansion.  In Sec.~\ref{linVd} we study the linear $V_d$
of Eq.~(\ref{Vdnf}).  We establish that a stable classical ground state exists,
and derive expressions for the masses of the pion and of the dilatonic meson.
In Sec.~\ref{matchT} we study the dilatation current and its anomalous divergence
at the level of the leading-order effective theory.
In Sec.~\ref{c1lim} we discuss the limit $n_f\nearrow n_f^*$,
where the ``phase boundary'' between chirally broken and infrared-conformal
theories is approached from within the chirally broken phase.
Finally, in Sec.~\ref{test} we return to the results of Sec.~\ref{linVd}, and give
a more hands-on description of how these predictions
can be tested within a given model.

%%%%%%%%
%\newpage
\subsection{\label{constVd} Constant $V_d$}
%%%%%%%%
An unusual feature of the classical potential~(\ref{Vtaueff})
is that it contains both an $O(1)$ term, and terms which are $O(\d)$.
The $\tc_{00}$ term in $V_d$ is $O(1)$,
while the terms linear in $n_f-n_f^*$ or in $m$ are $O(\d)$ in
the power counting of Sec.~\ref{conj}.  If we consider the $O(1)$ part
all by itself, the classical potential is $U(\t) = \tc_{00}\, e^{4\t}$.
Boundedness from below then requires $\tc_{00}>0$,
for which the classical vacuum
corresponds to $\svev{\t}\to -\infty$, or, equivalently, $\svev{e^\t}=0$.

In order to better understand this behavior, let us take into account
also the mass term, while still neglecting
the linear term in $V_d$.  In order to avoid unboundedness below
we must now also require $0<y<4$ (compare Eq.~(\ref{yrange})).
The classical potential then has a unique minimum $\svev{\t}=v$
given by
\begin{equation}
  e^{(4-y)v} = \frac{ym}{4\tc_{00}\cm} \ .
\label{solvec0}
\end{equation}
Now $v$ is finite for $m>0$, while in the chiral limit $v\to -\infty$
is recovered.

The ``runaway'' solution~(\ref{solvec0}) is not a consistent starting
point for the low-energy expansion.
In the case at hand, the physical decay constants
are the rescaled decay constants in Eq.~(\ref{fphys}), except
that $v$ is the solution of Eq.~(\ref{solvec0}).
Since $e^v \propto m^{\frac{1}{4-y}}$, we see that
the physical decay constants vanish in the chiral limit,
which is unacceptable.
The pion decay constant must tend to a finite value in the chiral limit.
Likewise, we expect
$\hf_\t$ to remain finite, as there is no reason that the dilatonic meson would
cease to exist in the chiral limit.

Equation~(\ref{Vdnf}) would naively suggest that the term linear in $\t$
can be neglected in the range $|n_f-n_f^*| \ll m/\L_{IR}$.
But, as we have seen, the outcome is inconsistent.
In the next subsection, we turn our attention to the role of
the linear term in $V_d$.

%%%%%%%%
%\newpage
\subsection{\label{linVd} Linear $V_d$}
%%%%%%%%
In this subsection we show that the linear $V_d$ of Eq.~(\ref{Vdnf}) gives rise
to a well-behaved classical solution.
Since we will be considering a fixed theory,
the $n_f$ dependence of $c_0$ and $c_1$ does not come into play.
We will revisit the role of this $n_f$ dependence in Sec.~\ref{c1lim} below.

We begin by considering the classical potential for $m=0$, where
\begin{equation}
  U(\t) = e^{4\t} (c_0 + c_1 \t) \ .
\label{UVlin}
\end{equation}
This potential is bounded from below provided that
\begin{equation}
  c_1 > 0 \ .
\label{c1}
\end{equation}
The saddle-point equation has a unique solution, the global minimum,
given by
\begin{equation}
  v = -1/4 - c_0/c_1 \ .
\label{minV}
\end{equation}
Although we are considering here a theory with fixed $n_f$,
the question arises
whether something might still go wrong if the selected theory
corresponds to a very small $n_f-n_f^*$, because,
according to Eq.~(\ref{minV}), $|v|$ will diverge if we keep selecting
theories with ever smaller $n_f-n_f^*$.  We will now show that
this is not the case.

Consider shifting the dilatonic meson field,
\begin{equation}
  \t \to \t + \D \ .
\label{shiftt}
\end{equation}
A glance at Eq.~(\ref{Vdnf}) reveals that this field redefinition induces
a corresponding redefinition of $c_0$,
\begin{equation}
  c_0 \to c_0 + c_1 \D \ .
\label{shiftc0}
\end{equation}
We can use this ``gauge freedom'' to set $c_0$ to any value we like,
and we will thus set
\begin{equation}
  c_0 = -c_1/4 \ .
\label{c01}
\end{equation}
This corresponds to shifting the $\t$ field by $\D=v$, where $v$ is
given by Eq.~(\ref{minV}), so that, after the shift,
the classical solution becomes
\begin{equation}
  v = 0 \ .
\label{v0}
\end{equation}
The remaining dependence of the tree-level lagrangian~(\ref{Leff})
on $\D$ is absorbed into a redefinition of the decay constants
$f_\p$ and $f_\t$, and the low-energy constants $B_\p$ and $B_\t$.
Since $\D$ is by construction equal to the original value of $v$
in Eq.~(\ref{minV}), the values of the redefined $f$'s and the $B$'s
are those given in Eqs.~(\ref{fphys}) and~(\ref{Bphys}).
Unlike in Sec.~\ref{constVd}, here we are working in the chiral limit,
and the expansion around the chiral limit is therefore well behaved.

It is easy to check that the solution $v=0$ is stable.
Indeed, if we add higher-order terms, $V_d$ generalizes to
\begin{equation}
  V_d = c_1(\t-1/4) + \t^2 f(\t) \ ,
\label{Vdquad}
\end{equation}
where we have used Eq.~(\ref{c01}).  According to Sec.~\ref{match},
$f(\t)=O((n_f-n_f^*)^2)$, making it a small correction for any given
theory for which $|n_f-n_f^*|\ll 1$.  First, one can easily check that $v=0$
remains a saddle point for any $f(\t)$.  Moreover, since
$|f(\t)|\ll c_1$, it follows that $v=0$ always remains a minimum.
The shifted classical solution $v=0$ is therefore a good starting point
for the diagrammatic low-energy expansion.

We next consider the tree-level masses.  In the chiral limit, the pion
is massless.  The tree-level mass $m_\t$ of the dilatonic meson
is obtained from the second derivative of $U(\t)$ evaluated
at the saddle point,
\begin{equation}
  m_\t^2 = 4 c_1 \hB_\t \ .
\label{mtau}
\end{equation}
In large-$N$ counting, all meson masses are $O(1)$.  This implies
that the low-energy constant $\hB_\t$ is $O(1)$ in large-$N$ counting
(a similar reasoning applies to $\hB_\p$).  Equation~(\ref{mtau}) also
implies that $m_\t \sim |n_f-n_f^*|^{1/2} \sim \d^{1/2}$,
which resembles the familiar behavior of the pion mass
in ordinary chiral perturbation theory, $m_\p\sim m^{1/2}$.

\medskip

Proceeding to the case $m>0$, we cannot solve for $v$ in closed form
any more.  The classical vacuum is now the solution of
\begin{equation}
  \frac{y m}{\cm c_1} = h(v) \equiv 4v e^{(4-y)v} \ .
\label{solvesdl}
\end{equation}
We keep $c_0 = -c_1/4$ as in Eq.~(\ref{c01}).  $\cm$ is now defined
in terms of the rescaled low-energy parameters of Eqs.~(\ref{fphys})
and~(\ref{Bphys}).  In order for the classical potential~(\ref{Vtaueff})
to remain bounded below for $m>0$, we need $0<y<4$.
As follows from Eq.~(\ref{yrange}), the actual value of $y$ appears to be
comfortably within this range.
Since both $c_1$ and $m$ are generically $O(\d)$, the left-hand side
of Eq.~(\ref{solvesdl}) is $O(1)$, and so is the solution $v$.
The function $h(v)$ is positive if and only if $v$ is.
Furthermore, for $v>0$, $h(v)$ is a monotonically increasing function.
Since we are assuming $m>0$, it follows that $v$ itself is
a monotonically increasing function of $m$.

Taking into account that the kinetic terms now get rescaled by $e^{2v}$,
the tree-level dilatonic meson mass is given by
\begin{equation}
  m_\t^2 = 4c_1\hB_\t e^{2v} (1+(4-y)v) \ .
\label{mintreetau}
\end{equation}
It is easy to see that $m_\t$ increases monotonically with $v$,
and thus with $m$.  The tree-level pion mass is given by
\begin{equation}
  m_\p^2 = 2 \hB_\p m e^{(y-2)v} = \frac{8 c_1\cm \hB_\p}{y}\, v e^{2v} \ .
\label{mintreepion}
\end{equation}
From the rightmost expression we learn that $m_\p$, too,
is monotonically increasing with $v$, and thus with $m$.

If we consider the limit where $m/(\cm c_1) \ll 1$, we can solve Eq.~(\ref{solvesdl})
to first order in this ratio, obtaining
\begin{equation}
  v = \frac{y m}{4\cm c_1} \ .
\label{dv}
\end{equation}
The correction to the dilatonic meson mass is
\begin{equation}
  \frac{m_\t^2}{m_\t^2(m=0)} = 1 + (6-y) v \ ,
\label{mtausqx}
\end{equation}
whereas the pion mass is given by
\begin{equation}
  m_\p^2 = 2\hB_\p m \ .
\label{mpisqx}
\end{equation}
This is the familiar result from ordinary chiral perturbation theory.
However, when $m/\cm\sim c_1$, the leading-order dependence of $m_\p^2$
on the quark mass, given by Eq.~(\ref{mintreepion}), is more complicated,
and not linear.

%%%%%%%%
%\newpage
\subsection{\label{matchT} Dilatation current in the leading-order effective theory}
%%%%%%%%
In this subsection we explore the basic properties of the dilatation current
and its divergence to leading order in the effective theory.
The actual derivation of the dilatation current is relegated to App.~\ref{SLO}.
The result can be expressed as (see Eq.~(\ref{LOcrnt}))
\begin{equation}
  S_\m = x_\n \Th_{\m\n} = x_\n (T_{\m\n} + K_{\m\n}/3) \ ,
\label{SLOagain}
\end{equation}
where $T_{\m\n}$ and $\Th_{\m\n}$ are, respectively,
the canonical and improved energy-momentum tensors
of the effective theory at the leading order.\footnote{%
  Since the fields of the effective theory are Lorentz scalars,
  the existence of an improvement term is to be expected \cite{CCJ}.
}

As expected, $S_\m$ contains a term linear in the dilatonic meson field.
Its origin is in the improvement term
\begin{eqnarray}
  K_{\m\n} &=& \frac{f_\t^2}{2}
  ( \d_{\m\n} \bo - \partial_\m \partial_\n ) e^{2\t}
\label{Kmn}\\
  &=& f_\t e^v ( \d_{\m\n} \bo - \partial_\m \partial_\n ) \tilt + \cdots \ .
\nonumber
\end{eqnarray}
where on the second line we have used Eq.~(\ref{cnnprop}).
Plugging this into Eq.~(\ref{SLOagain}) gives
\begin{equation}
  S_\m = f_\t e^v \partial_\m \tilt + \cdots \ ,
\label{Slin}
\end{equation}
showing that the leading-order decay constant of the dilatonic meson
is $f_\t e^v$.  [Note that, according to Eq.~(\ref{dLa}), the off-shell
divergence $\partial_\m S_\m$ contains
$\partial_\m (x_\n K_{\m\n}/3) = (\partial_\m x_\n) K_{\m\n}/3
= K_{\m\m}/3 = (f_\t^2/2) \bo e^{2\t}$, which is equivalent to Eq.~(\ref{Slin}).]

The anomalous conservation equation may be read off from Eqs.~(\ref{dL})
and~(\ref{Tmn}).  Using also Eqs.~(\ref{gammay}) and~(\ref{gammastar}),
it takes the form
\begin{equation}
  \partial_\m S_\m = c_1 f_\t^2 B_\t e^{4\t}
  +(1+\g_m^*)\frac{f_\p^2 B_\p m}{2}\, e^{y\t} \tr(\S+\S^\dagger) \ .
\label{LOnonc}
\end{equation}
If we now introduce
\begin{equation}
  \bj\j({\rm EFT})
  \equiv \frac{\partial \cl}{\partial \cs}
  = -\frac{f_\p^2 B_\p}{2}\, e^{y\t} \tr (\S+\S^\dagger) \ ,
\label{eftbjj}
\end{equation}
with $\cl$ given by Eq.~(\ref{Leff}), we may rewrite Eq.~(\ref{LOnonc}) as
\begin{equation}
  \partial_\m S_\m = c_1 f_\t^2 B_\t e^{4\t} - (1+\g_m^*) m \bj\j({\rm EFT}) \ .
\label{noncop}
\end{equation}
A comparison to the microscopic theory (Sec.~\ref{dST}) leads
to the identification
\begin{equation}
  \tb F^2({\rm EFT}) = -c_1 f_\t^2 B_\t e^{4\t} \ .
\label{eftFF}
\end{equation}
If we take the expectation value of Eq.~(\ref{eftFF}),
and demand consistency with the microscopic theory, we obtain
\begin{equation}
  \tb \vev{F^2} = -c_1 f_\t^2 B_\t e^{4v} \ .
\label{svevFF}
\end{equation}
Since in the effective theory we were led to require that $c_1>0$, and
we have that $\tb<0$, it follows that $\svev{F^2}>0$, for $n_f\,\ltap\, n_f^*$.
We recall that $F^2$ can mix with the identity operator.
Different prescriptions to subtract the identity part of $\svev{F^2}$
can vary,
and, as a result, even the sign of the subtracted $\vev{F^2}$ is not
uniquely defined.  Interestingly, the effective theory appears to provide
a natural framework to define the (subtracted) expectation value $\svev{F^2}$,
which, as we have seen, implies that it is positive.

Finally, another interesting consequence
is that, in the chiral limit, we may use Eq.~(\ref{mtau})
(as well as~(\ref{fphys}) and~(\ref{Bphys})) to rewrite Eq.~(\ref{svevFF}) as
\begin{equation}
  -4\tb \vev{F^2} = \hf_\t^2 m_\t^2 \ .
\label{GMORtau}
\end{equation}
This result closely resembles the GMOR relation\footnote{%
  Note that $\bj\j$ is summed over flavors.
}
\begin{equation}
  -\frac{2m}{N_f}\,\svev{\bj\j} = \hf_\p^2 m_\p^2 \ ,
\label{GMOR}
\end{equation}
which is reproduced using Eqs.~(\ref{mpisqx}) and~(\ref{eftbjj}).
The existence of a GMOR-like relation for the dilatonic meson
is not entirely obvious.  The original GMOR relation is derived from
the partial conservation of the axial current by making use of the fact
that the pion pole is isolated from the 3-pion cut.
By contrast, for any finite $N_c$ and $N_f$, we expect that the mass
of the dilatonic meson will remain non-zero in the chiral limit.
Therefore, for $m=0$ the dilatonic meson is unstable, and can decay
into two pions.  The $\t\to \p\p$ vertex comes from expanding
the pion kinetic term~(\ref{Lp}), and is given by
$\sim f_\t^{-1}\, \t \tr(\partial_\m\p \partial_\m\p)$,
showing that $f_\t^{-1}$ plays
the role of the coupling constant.  The resulting decay rate satisfies
\begin{equation}
  \frac{\G_\t}{m_\t} \sim \frac{m_\t^2 N_f^2}{\hf_\t^2}
  \sim \frac{m_\t^2}{\L_{IR}^2} \sim |n_f-n_f^*| \ ,
\label{decay}
\end{equation}
where in the first equality we have used that the number of pions
is $\sim N_f^2$.  We see that, thanks to the derivative interaction
of the pions in the chiral limit, the ratio $\G_\t/m_\t$
is parametrically small, and so the dilatonic meson is a narrow resonance
for $n_f$ close to $n_f^*$.

%%%%%%%%
%\newpage
\subsection{\label{c1lim} The limit $n_f\nearrow n_f^*$}
%%%%%%%%
In the previous subsections we have studied the tree-level structure
of a single theory with specific values of $N_c$ and $N_f$,
and, thus, a fixed value of $n_f$.
In this subsection, we explore the limit $n_f\nearrow n_f^*$,
which necessarily involves the comparison of different theories.

Having in mind that we will now be varying $n_f$ requires us to
revisit two elements of the discussion of Sec.~\ref{linVd}.  First,
reinstating the $n_f$ dependence (and remembering that we always
consider $n_f<n_f^*)$
the bound~(\ref{c1}) reads
\begin{equation}
  \tc_{11} < 0 \ .
\label{tc11}
\end{equation}
Second, in Sec.~\ref{linVd},
the ``gauge freedom'' of shifting the $\t$ field has
allowed us to set its expectation value to zero in the chiral limit.
In order to facilitate the comparison of different theories
we may still shift the $\t$ field,
but the shift must now be uniform across these theories, that is, independent of $n_f$.
Parametrically, this shift is $O(1)$, so that
it does not mix different orders in the expansion in $n_f-n_f^*$.
Comparing Eqs.~(\ref{Vdnf}) and~(\ref{shiftc0}),
we see that the effect of an allowed shift is now
\begin{subequations}
\label{tcshift}
\begin{eqnarray}
  \tc_{00} &\to& \tc_{00} \ ,
\label{tcshifta}\\
  \tc_{01} &\to& \tc_{01} + \tc_{11} \D \ .
\label{tcshiftb}
\end{eqnarray}
\end{subequations}
It follows that the value of the low-energy constant $\tc_{01}$ is arbitrary.
We will henceforth use the freedom
to perform a uniform shift to set $\tc_{01}=0$.
By contrast, the leading-order coefficient $\tc_{00}$ is invariant
under the shift, and thus, it can have physical consequences.

The role of $\tc_{00}$ is best illustrated by considering a concrete example.
Starting from the rescalings in Eqs.~(\ref{fphys})
and~(\ref{Bphys}), we consider the dimensionless ratio\footnote{%
  Here we are, in effect, using the reference infrared scale
  $\L_{IR} = 4\p \hf_\p N_c^{-1/2}$ (compare Eq.~(\ref{LIR})).
  The ratio~(\ref{Bf}) allows us to track
  the $n_f$ dependence at fixed $N$.  The $N$-dependence can easily be
  divided out for the purpose of comparing theories with different $N$.
}
\begin{equation}
  \frac{m\hB_\p}{\hf_\p^2}
  = \frac{mB_\p}{f_\p^2}\, e^{-(1+\g_m^*) v} \ ,
\label{Bf}
\end{equation}
where we have used Eqs.~(\ref{gammay}) and~(\ref{gammastar})
to trade $y$ with the mass anomalous dimension
$\g_m^*$ at the sill of the conformal window, Eq.~(\ref{gammastar}).
The left-hand side of Eq.~(\ref{Bf}) can be
determined in terms of the pion mass close to the chiral limit
and the (physical) pion decay constant
\begin{equation}
  \frac{m\hB_\p}{\hf_\p^2}
  = \frac{m_\p^2}{2\hf_\p^2} \ ,
\label{Bfcond}
\end{equation}
where we have used Eq.~(\ref{mpisqx}).
By plugging Eq.~(\ref{Vdnf}) into Eq.~(\ref{minV}), we reexpress the
classical solution as
\begin{equation}
  v = -\frac{1}{4} -\frac{\tc_{00}}{\tc_{11}(n_f-n_f^*)} \ ,
\label{minVnf}
\end{equation}
where we have used the new ``gauge choice'' $\tc_{01}=0$.
We see that the dimensionless ratio $m_\p^2/\hf_\p^2$ depends
(exponentially) on $\tc_{00}/\tc_{11}$.
If we measure this ratio for sufficiently many walking theories
(and close enough to the chiral limit),
we will be able to extract from it the value of
\begin{equation}
  \frac{\tc_{00}}{\tc_{11}} = \frac{B_\t\tc_{00}}{B_\t\tc_{11}} \ .
\label{c0by1}
\end{equation}
A linearly independent relation is obtained by measuring
the mass of the dilatonic meson, which, in the present context,
becomes (compare Eq.~(\ref{mtau}))
\begin{equation}
  m_\t^2 = 4 \tc_{11}(n_f-n_f^*) B_\t e^{2v} \ .
\label{mtaugen}
\end{equation}
Using both measurements for a series of theories
in the limit of small $n_f-n_f^*$ and $m$, we should thus be able
to separately determine $\tc_{00}B_\t$ and $\tc_{11}B_\t$.

A technical comment is that, since the leading-order lagrangian depends on
$B_\t V_d$, we only have access to products
such as $c_n B_\t$ or $\tc_{nk} B_\t$.
This behavior is familiar from
the standard chiral lagrangian, where only the product $m B_\p$
has an invariant meaning.
In both cases, the separate determination of the $B$ parameter
requires an additional extraneous prescription.

Returning to Eq.~(\ref{minVnf}) we find in the limit $n_f\nearrow n_f^*$
the asymptotic behavior
\begin{equation}
  v \to \Bigg\{ \begin{array}{ll}
  -\infty \ , & \qquad \tc_{00}>0 \ , \\
  +\infty \ , & \qquad \tc_{00}<0 \ .
  \end{array}
\label{limv}
\end{equation}
We conclude that $\hf_\p(n_f)$, $\hf_\t(n_f)$ and $\hB_\t(n_f)$ vanish
in this limit when $\tc_{00}>0$,
or, alternatively, blow up when $\tc_{00}<0$.\footnote{%
  The same conclusion applies to $\hB_\p(n_f)$ provided that $\g_m^*<1$.
}
Intuitively, the correct behavior is that all dimensionful low-energy parameters
will vanish as the conformal window is approached.  This would imply that
the ``phase boundary'' between chirally broken and infrared-conformal theories
is smooth.  We thus speculate that the correct sign is $\tc_{00}>0$.
However, this is merely a speculation.
The reason is that when we look at, \eg, the $n_f$ dependence of $\hf_\p$,
we are comparing the pion decay constants of different theories.
Unless we have a common dimensionful reference scale,
such a comparison is meaningless.  Taking a low-energy parameter,
such as $f_\p$, as the common reference scale would only work if we have
a way to determine $f_\p$ itself,
and not just $\hf_\p$, from the microscopic theory.
This requires insight into the microscopic theory
not provided by the low-energy effective theory.

In order to avoid confusion, we note that in Sec.~\ref{constVd}
we have considered a fixed theory, and the conclusion that the
classical solution~(\ref{solvec0}) is not a consistent starting point followed
from the requirement that the chiral limit of that particular theory must not
be singular.  Once we take into account the linear term in $V_d$,
the sign of $\tc_{00}$ becomes unconstrained, and we need instead
the constraint~(\ref{c1}), or, equivalently,~(\ref{tc11}).

The description in terms of the effective low-energy theory
developed in this paper necessarily breaks down inside the conformal window.
The reason is that meson states no longer exist in the chiral limit,
where, instead, all correlation functions obey power laws
with non-trivial exponents.  In view of Eq.~(\ref{tc11}), the product
$c_1= \tc_{11}(n_f- n_f^*)$ will be negative for $n_f>n_f^*$, and, as a result,
the tree-level potential will become unbounded from below.
It is reassuring to see that, through this classical behavior,
the low-energy theory ``knows''
that it must cease to be valid as the conformal window is entered.

To conclude this section, we consider another dimensionless ratio, which,
through the GMOR relation, directly influences the pion mass.
This is the value of the flavor-singlet fermion condensate,
expressed in units of the pion decay constant,
\begin{equation}
  \frac{\vev{\bj\j}}{\hf_\p^3}
  = -
  \frac{B_\p}{f_\p}\, e^{-\g_m^* v} \ .
\label{condscale}
\end{equation}
If indeed $\tc_{00}$ is positive (and since $\g_m^*$ is expected to be
positive, too), we have that $e^{-\g_m^* v}>1$.  This result then predicts
a condensate enhancement that depends on the mass anomalous dimension,
as well as, through Eq.~(\ref{minVnf}), on the low-energy constants $\tc_{00}$
and $\tc_{11}$.  While condensate enhancement was predicted long ago
within the gap-equation treatment \cite{BH,YBM,AY,AKW,BG,CG}, here we obtain it
as a quantitative prediction
within the framework of a systematic, low-energy expansion,
provided that $\tc_{00}>0$.

%%%%%%%%
%\newpage
\subsection{\label{test} Testing the effective theory}
%%%%%%%%
In this section we have studied several tree-level predictions
of the effective theory, first for a given model, and then
as the limit $n_f\nearrow n_f^*$ is approached.  In practice,
the ability to test the effective theory by collecting data from
many models is limited by its cost.  Thus,
in this subsection we return to a specific model
with fixed $N_c$ and $N_f$, and discuss in some more detail how
the effective theory can be put to a test in this framework.

We will focus on the dependence
of simple observables on the fermion mass $m$.  Of course, virtually any
physical observable of the low-energy sector can be used to test the
predictions of the effective theory.  The discussion below is only meant
as an illustration, providing a more hands-on view on the results
already presented in Sec.~\ref{linVd}.

We first discuss the determination of the exponent $y$,
or, equivalently, the mass anomalous dimension $\g_m^*$ at the sill
of the conformal window (Eqs.~(\ref{gammay}) and~(\ref{gammastar})).
The results of Sec.~\ref{LOV}, which are based on
our main dynamical assumption~(\ref{tbg}), imply that for a given model,
the mass anomalous dimension at the dynamical infrared scale
$\g_m(\L_{IR})$ can be expanded as
\begin{equation}
  \g_m(\L_{IR}) = \g_m^* + c_{\g,1} (n_f-n_f^*) + \cdots \ .
\label{expandgm}
\end{equation}
In comparison with Eq.~(\ref{tbg}), the main difference is that
at the scale $\L_{IR}$, the trace anomaly
vanishes with $n_f-n_f^*$, whereas the mass anomalous dimension stays finite.
It then follows that we may reexpress $e^{y\t}$ in Eq.~(\ref{Lm}) as
\begin{equation}
  e^{y\t} = e^{(3-\g_m^*)\t}
  = e^{(3-\g_m(\L_{IR}))\t} (1 +  c_{\g,1} (n_f-n_f^*)\t + \cdots) \ .
\label{expandy}
\end{equation}
We conclude that the discrepancy arising from using $\g_m(\L_{IR})$ of the model
with fixed $N_f$ and $N_c$, instead of $\g_m^*$,
will be absorbed at the next-to-leading order by adjusting the
coefficient $\tc_{M,11}$ occurring
in the expansion of $V_M$ (see Eq.~(\ref{LmV}) and Sec.~\ref{lagnlo} below).

With an estimate for $y$ at hand, we may now use the numerical results
for the pion mass $m_\p$, measured at several different values of
the fermion mass $m$, and substitute them into Eq.~(\ref{mintreepion})
in order to determine the functional form of $v(m)$
together with the low-energy constant $\hB_\p$.  The latter may be found
by considering the small-$m$ behavior, because,
by construction, the shifted dilaton expectation value of Sec.~\ref{linVd}
vanishes in the chiral limit.

At this point we may carry out two tests of the effective theory.
First, the function $h(v)$, defined in Eq.~(\ref{solvesdl}),
must depend linearly on $m$ once we substitute $v=v(m)$.
Another test is that the $m$-dependence
of the dilatonic meson mass should agree with Eq.~(\ref{mintreetau}),
which, in turn, allows us to determine the product $c_1 \hB_\t$.

A word of caution is that, since we are using the predictions
of the leading-order lagrangian, the results are only expected to provide
a good description of the low-energy sector when all the expansion parameters,
$m/\L_{IR}$, $1/N_c$ and $n_f-n_f^*$, are sufficiently small.

As we have explained, the effective theory is restricted to theories with
fermions in the fundamental representation because only in this case can we
turn the ratio $N_f/N_c$ into an effectively continuous parameter by invoking
the Veneziano limit.  While this would necessarily be of a speculative nature,
let us give a thought to
models with higher-representation fermions (such as the $SU(3)$-sextet
model studied in Refs.~\cite{BMWs,BMW6}).  Consider the euclidean
partition function defined by varying {\it continuously} the number of
flavors $N_f$.  While the theory is non-local whenever $N_f$ is not
an integer, it is a well-defined statistical mechanics system.
That system might become infrared conformal
at some (non-integer!) value of $N_f$, which we denote as $N_f^*$.
In the fortuitous event that there exist an integer $N_f$ such that,
first, $N_f<N_f^*$ so that the $N_f$-flavor theory undergoes spontaneous
chiral symmetry breaking, and, second,
$N_f^*-N_f\ll 1$, the low-energy sector of the $N_f$-flavor
higher-representation theory might
admit an expansion in $m/\L_{IR}$ and $N_f-N_f^*$.

Testing this scenario can be done along similar lines.
For the actual effective field theory framework
(with fermions in the fundamental representation),
the success of the low-energy description at a given order
depends on the smallness of the expansion parameters $1/N_c$ and $n_f-n_f^*$
for the model under consideration; the fermion mass can always be dialed
such that $m/\L_{IR}$ is small enough.
Similar tests of the low-energy sector can be carried out in any model with
higher-representation fermions and a light scalar meson,
and the hope would be that that model
is well-described by an expansion in $N_f-N_f^*$.

%%%%%%%%
%\newpage
\section{\label{NLO} Effective field theory at one loop}
%%%%%%%%
In this section we study the effective theory at the next-to-leading order
(NLO).  In Sec.~\ref{lagnlo} we write down the NLO lagrangian.  Then,
as an example, we work out in Sec.~\ref{Veff} the one-loop effective potential
from a dilatonic meson loop, and confirm that the
NLO lagrangian contains the necessary counterterms to renormalize it.

%%%%%%%%
%\newpage
\subsection{\label{lagnlo} Next to leading order lagrangian}
%%%%%%%%
Like the standard chiral lagrangian, our low-energy expansion involves
a derivative expansion, as well as an expansion in the fermion mass $m$,
or, more generally, in the chiral source $\c$.
As we will shortly see, the systematic expansion in $\partial_\m$ and $\c$
provides the main organizing principle for the successively higher orders
in the effective theory.

Our setup contains two additional
small parameters: $1/N$, and $n_f-n_f^*$.  The systematic expansion
in $n_f-n_f^*$ is realized in a very simple way: each operator
that we encounter while expanding in $\partial_\m$ and $\c$
will be multiplied by its own invariant potential $V_I(\t-\s)$,
which, after setting $\s=0$, reduces to $V_I(\t)$.
The invariance property of these potentials under
the dilatation~(\ref{scleffb}), namely, the transformation rule
$V_I(\t(x))\to V_I(\t(\l x))$, is maintained by promoting the
power-series coefficients of the potentials to (global) spurions,
whose transformation properties are derived in App.~\ref{diltc}.
At each order in the expansion defined by the power counting
of Sec.~\ref{conj} the potentials $V_I(\t)$ can be truncated
consistent with that order, and the effective theory thus contains a
finite number of low-energy constants.

As for $1/N$, the reason why we invoke the Veneziano limit is to justify the
treatment of $n_f-n_f^*$ as a continuous parameter.
The only explicit $N$-dependence is that the $\S$ field takes values
in $SU(N_f)$.  In addition, the decay constants scale with $N$,
as discussed in Sec.~\ref{conj}.  The parameter $1/N$ does not play any
direct role in organizing the effective lagrangian.

Before we turn to the construction of the NLO lagrangian,
let us reconsider the freedom to shift the $\t$ field
and its effect on the LO lagrangian.
In order to develop the low-energy expansion of a given theory,
in this section we will choose to
shift the $\t$ field by $\D=-\tc_{00}/(\tc_{11}(n_f-n_f^*))$.
This shift sets $\tc_{00}=0$, thereby eliminating the only $O(1)$ term
present in the LO lagrangian, and the entire LO lagrangian
becomes $O(\d)$.  This new ``gauge choice'' leaves $\tc_{01}$ intact.
Recall that, in Sec.~\ref{c1lim}, we made use of an $n_f$-independent shift
to argue that the value of $\tc_{01}$ is unphysical.
Nevertheless, here it is advantageous to keep it free,
because this manifestly maintains dilatation invariance of
the $O(n_f-n_f^*)$ part of $V_d$ (see Eq.~(\ref{Vdnf}) and App.~\ref{diltc}).

The classical vacuum is now given by $v=-1/4-\tc_{01}/\tc_{11}$,
which is $O(1)$, and hence any factors of $e^v$ which arise from the expansion
of the $\t$ field, Eq.~(\ref{cnnprop}), are $O(1)$ too.\footnote{%
  This reduces to the solution $v=0$ of Eq.~(\ref{v0})
  in the special case that $\tc_{01}=-\tc_{11}/4$.
}
The classical solution will now be modified by higher-order terms
of the kind studied in Eq.~(\ref{Vdquad}).  But the corresponding change
in $v$ will be $O(n_f-n_f^*)$, so that the systematic nature of
the low-energy expansion is respected.

\medskip

The NLO lagrangian consists of four kinds of operators.  The first kind
comes from picking the $O(\d^2)$ terms from the LO lagrangian of Eq.~(\ref{LeffL})
while setting $\s=0$.
This amounts to picking the terms $(n_f-n_f^*)(\tc_{I,01}+\tc_{I,11}\t)$
from the expansions of $V_M$, $V_\p$ and $V_\t$, because these potentials
multiply operators that are $O(\d)$ by themselves.
The role of $V_d$ is again special, because $\tcl_d$ is the only part of the
effective theory which involves no small parameters other than $n_f-n_f^*$.
The $O(\d^2)$ terms coming from $V_d$ give rise to the NLO operators
\begin{subequations}
\label{Vd2nd}
\begin{eqnarray}
  Q_1^d &=&  \tc_{02}\, (n_f-n_f^*)^2 f_\t^2 B_\t \, e^{4\t} \ ,
\label{Vd2nda}\\
  Q_2^d &=&  \tc_{12} \t \, (n_f-n_f^*)^2 f_\t^2 B_\t \, e^{4\t} \ ,
\label{Vd2ndb}\\
  Q_3^d &=&  \tc_{22} (\t^2/2) (n_f-n_f^*)^2 f_\t^2 B_\t \, e^{4\t} \ .
\label{Vd2ndc}
\end{eqnarray}
\end{subequations}
The dilatation transformation properties of the low-energy constants
$\tc_{nk}$, given in App.~\ref{diltc}, ensure that the spacetime integral of
$Q_1^d+Q_2^d+Q_3^d$ is invariant.

The remaining terms in the NLO lagrangian arise from the expansion
in $\partial_\m$ and $\c$, which we will refer to as the ``chiral expansion.''
Since these terms are already $O(\d^2)$ on account of their dependence
on derivatives and on the chiral source $\c$,
it follows that the invariant potentials
multiplying all these terms may be set equal to 1.
At the next-to-next to leading order (NNLO), we will need the
$O((n_f-n_f^*)^3)$ part of $V_d$, the $O((n_f-n_f^*)^2)$ parts of $V_M$,
$V_\p$ and $V_\t$, and the $O(n_f-n_f^*)$ parts of invariant potentials that
multiply NLO operators arising from the chiral expansion.

The first kind of chiral-expansion NLO operators
comes in one-to-one correspondence
with the NLO operators of standard chiral perturbation theory \cite{GL2}.
Each NLO operator
\begin{equation}
  \tQ_i = \tQ_i(\S,\c,\partial_\m) \ ,
\label{Qtilde}
\end{equation}
appears in the lagrangian in the form
\begin{equation}
  Q_i^\p = e^{4\t}\, \tQ_i(\S,e^{(y-4)\t}\c,e^{-\t}\partial_\m) \ .
\label{Qexp4t}
\end{equation}
The NLO action constructed from the operators $Q_i^\p$ is dilatation
invariant.
A second set of chiral-expansion NLO operators consists of operators that
depend on the dilatonic meson field only.  There are three of them,\footnote{%
  We omit the low-energy constants that multiply each operator
  in Eqs.~(\ref{Q}) and~(\ref{MX}).
}
\begin{subequations}
\label{Q}
\begin{eqnarray}
  Q^\t_1 &=& [(\partial_\mu \t)^2]^2 \ ,
\label{Q1}\\
  Q^\t_2 &=& (\bo \t)^2 \ ,
\label{Q2}\\
  Q^\t_3 &=& \bo \t (\partial_\mu \t)^2 \ .
\label{Q3}
\end{eqnarray}
\end{subequations}
Finally, there are operators that depend on both $\S$
and derivatives of $\t$,
\begin{subequations}
\label{MX}
\begin{eqnarray}
  Q^{\rm mix}_1 &=& \partial_\m \t \partial_\n \t
  \tr(\partial_\m \S^\dagger \partial_\n \S) \ ,
\label{MX1}\\
  Q^{\rm mix}_2 &=&
  (\partial_\m \t)^2  \tr(\partial_\n \S^\dagger \partial_\n \S) \ ,
\label{MX2}\\
  Q^{\rm mix}_3 &=& \bo \t \tr(\partial_\n \S^\dagger \partial_\n \S) \ ,
\label{MX3}\\
  Q^{\rm mix}_4 &=& e^{(y-2)\t}\, (\partial_\m \t)^2
  \tr\Big(\c^\dagger \S + \S^\dagger \c\Big) \ ,
\label{MX4}\\
  Q^{\rm mix}_5 &=&
  e^{(y-2)\t}\, \bo \t \tr\Big(\c^\dagger \S + \S^\dagger \c\Big) \ .
\label{MX5}
\end{eqnarray}
\end{subequations}

%%%%%%%%
%\newpage
\subsection{\label{Veff} Effective potential at NLO}
%%%%%%%%
In this subsection we calculate, as an example,
the one-loop effective potential
coming from a dilatonic meson field in the loop.\footnote{%
  We did not calculate the contribution to the effective potential
  coming from a pion loop, for which the power counting should work
  in the usual way.
}
As in the previous
subsection, we use the $\t$ shift to set $\tc_{00}=0$ while
leaving $\tc_{01}$ free.
The easiest way to obtain the effective potential is via the background field
method.  Instead of a constant classical vacuum $v$, we introduce
a background field $u(x)$, assumed to be slowly varying,
and replace Eq.~(\ref{cnnprop}) by
\begin{equation}
  \t(x) = u(x) + \tilt(x)/f_\t  \ ,
\label{splitq}
\end{equation}
where, as before, $\tilt(x)$ is the quantum part.\footnote{%
  Because the classical background is not constant,
  we normalize the quantum part only by the original $f_\t$.
}
We will be applying an adiabatic approximation, where derivatives
of the background field are neglected.
The result of the calculation is a functional of the background field $u(x)$,
in which we are eventually to replace $u(x)$ by the full field $\t(x)$.

We start by expanding $S$ to quadratic order in $\tilt$,
\begin{equation}
  S^{(2)} = \frac{\m^{d-4} f_\t^2}{2} \int d^dx\,
  \left( e^{2u}(\partial_\m\tilt)^2 + \cv''(u)\tilt^2 \right)\ ,
\label{Squad}
\end{equation}
where now (compare Eq.~(\ref{Vtaueff}))
\begin{equation}
  \cv(u) = B_\t V_d(u) e^{4u}
  - \frac{f_\p^2 B_\p}{2f_\t^2 } \, e^{yu}
  \tr\Big(\c^\dagger\S+\S^\dagger\c\Big) \ ,
\label{taupot}
\end{equation}
and, integrating over $\tilt$, we find
\begin{subequations}
\label{V1}
\begin{eqnarray}
  V_1 &=& \half\int\frac{d^dp}{(2\p)^d}\log\left( e^{2u}p^2 + \cv''(u)\right)
\label{V1a}\\
  &=& -\half\frac{\partial}{\partial\a}
  \int\frac{d^dp}{(2\p)^d}\left( e^{2u}p^2 + \cv''(u)\right)^{-\a}\Big|_{\a=0}
\label{V1b}\\
  &=& -\half\,\frac{\G(-\frac{d}{2})}{(4\p)^{d/2}}
  \left(e^{-2u} \cv''(u)\right)^{\frac{d}{2}}
\label{V1c}\\
  &=& -\frac{1}{64\p^2} \left(e^{-2u} \cv''(u)\right)^2
\NON
  && \rule{0ex}{3.5ex} \times \left(\frac{2}{4-d}-\g+\frac{3}{2}
  -\log\left(\frac{e^{-2u} \cv''(u)}{4\p\m^2}\right)
  +O(d-4) \right)\ .
\label{V1d}
\end{eqnarray}
\end{subequations}
We confirmed this result by a diagrammatic calculation, see App.~\ref{resumVeff}.
The effective potential is thus given by Eq.~(\ref{V1d}) with the replacement
$u(x)\to \t(x)$.

The divergence of the effective potential is proportional to
$(e^{-2\t} \cv''(\t))^2$.
It is straightforward to check that the counterterms needed to remove it
are a linear combination of the NLO operators of Sec.~\ref{lagnlo}.
As an example, if we set $\c=0$ then the required counterterms are
the three NLO operators in Eq.~(\ref{Vd2nd}).
This calculation provides a non-trivial check of the constraints
that follow from the power counting and from the symmetries
of the effective theory.

%%%%%%%%%%%%%%%%%%%%%%%%%%%
%\newpage
\section{\label{disc} Discussion}
%%%%%%%%%%%%%%%%%%%%%%%%%%%
We begin this concluding section with a summary
of the main technical ingredients of our work.
We have proposed a low-energy effective action that accommodates
dilatation symmetry into the framework of the chiral lagrangian.
Previously, the chiral lagrangian was extended to accommodate
the anomalous $U(1)_A$ symmetry.  What these two extensions have in common is
that the new symmetry, be it dilatations or $U(1)$ axial transformations,
is abelian.  In both cases, the dependence of the low-energy theory
on the singlet dynamical field associated with the abelian symmetry
is virtually unconstrained at the algebraic level.  For the low-energy
theory to have any predictive power, we must be able to establish
a power-counting hierarchy that will suppress interactions with
increasingly large number of singlet particles.

For the $U(1)_A$ symmetry, the needed hierarchy was established
in the usual large-$N_c$ limit, in which the axial anomaly vanishes \cite{GL2,KL}.
The low-energy expansion is then a double expansion: in the fermion mass $m$,
and in $1/N_c$.  In the case of dilatations, we are able to propose
a suitable starting point only for fermions in the fundamental representation.
The low-energy expansion involves a triple limit: the chiral limit $m\to 0$;
the Veneziano large-$N$ limit in which the ratio $n_f=N_f/N_c$
effectively becomes a continuous parameter; and the limit
in which $n_f$ approaches the sill of the conformal window $n_f^*$ from below.

The main hypothesis underlying our derivation is that, when we increase
both $N_c$ and $N_f$ such that $n_f$ tends to $n_f^*$ from within the
chirally broken phase, the trace of the energy-momentum tensor
in the chiral limit tends to zero when probed at the scale at which
spontaneous chiral symmetry breaking sets in.
This hypothesis is compatible with the notion of a smooth
``conformal phase transition,'' and with the existence of an
infrared fixed-point as the conformal window is entered.
While the low-energy theory is not directly
sensitive to the fixed-point value of the coupling, it depends explicitly
on the corresponding value of mass anomalous dimension.

For $n_f\nearrow n_f^*$, conservation of the dilatation current
in the chiral limit is in effect recovered.
However, instead of the dilatonic meson becoming a true Nambu-Goldstone
boson, this limit is at the edge of the region of validity of the effective
theory.   The effective theory breaks down through a diverging
vacuum expectation value $v=\svev{\t}$ at $n_f=n_f^*$,
leading to unphysical values of the decay constants $f_\p$ and $f_\t$.
For $n_f>n_f^*$, the classical potential of the effective theory
becomes unbounded from below.
This singular behavior suggests that pions cease to exist as the conformal
window is entered, consistent with the absence of chiral symmetry breaking
inside the conformal window.   It also suggests the absence of a true
Nambu-Goldstone boson for dilatation symmetry inside the conformal
window.   However, we emphasize that as the effective theory breaks down
at the conformal sill, it does not tell us anything about the physics
on the other side of the sill.

An external dilaton source $\s(x)$ which, in the microscopic theory,
is coupled to the trace anomaly, \ie, to the quantum part
of the operatorial trace of the energy-momentum tensor,
communicates information about the operatorial trace to the effective theory.
By matching $\s$-dependent correlation functions between the microscopic
and the effective theory we were able to establish the desired
power-counting hierarchy.  Successive powers of $\t-\s$ in the
expansion of the invariant potentials $V_I(\t-\s)$ are suppressed by
equal powers of $n_f-n_f^*$.

In comparison to the low-energy framework that includes the $U(1)_A$ symmetry,
a simplification that occurs in our case is that all relevant currents
are renormalization-group invariant.  This includes the energy-momentum tensor,
the dilatation current, and the non-singlet axial currents.
A related choice we have made is that
we opted to avoid the introduction of external gauge fields
of any kind.  In particular, we derived the dilatation current
in the effective theory directly using the Noether procedure.

In the microscopic theory, we have avoided multiple operator insertions
at coinciding points.  As a result,
we did not probe any $c$-number trace anomalies.
In flat space, these $c$-number anomalies reduce to four-derivative terms.
Whether the four-derivative terms of the
next-to-leading order lagrangian may be constrained via matching conditions
similar to those of Refs.~\cite{ST,KS} we leave as an open question.
Studying this question requires the
promotion of the global dilatation symmetry to local Weyl invariance,
which involves the introduction of the metric tensor \cite{DL}, and perhaps
also Weyl gauging \cite{Weylgg}.

The anomalous conservation law of the dilatation current in the
effective theory suggests that, up to a suitable proportionality constant,
the field strength operator of the microscopic theory, $F^2$,
is represented in the low-energy theory by $e^{4\t}$, where $\t$
is the dilatonic meson field.  Now, the classical potential
of the effective theory is $e^{4\t}(c_0+c_1\t)$ in the chiral limit.
If we make use of the above identification,
it takes the form $F^2(c_0+(c_1/4)\log F^2)$.  This effective lagrangian was
proposed long ago as a phenomenological lagrangian for Yang-Mills theory
\cite{JS,MS}.  The difference is that, in our case, the classical potential
arises in an effective theory that provides a systematic low-energy expansion
of gauge theories near (but below) the bottom of the conformal window.

Among the notable predictions of the tree-level effective theory
is the result that the masses of both the dilatonic meson and the pion
are monotonically increasing with the fermion mass $m$.
In the region where $m_\p\sim m_\t$, the dependence of the pion mass on $m$
is more complicated than the standard GMOR relation.
We have also derived a GMOR-like relation
for the dilatonic meson in the chiral limit.
The reason why such a relation exists, in spite of the fact
that the dilatonic meson can decay into two pions, is that the pions
have a derivative coupling in the chiral limit, and, as a result,
the ratio of the dilatonic meson's decay rate to its mass tends to zero
when $n_f$ tends to $n_f^*$.
Finally, we obtained a quantitative prediction for the enhancement
of the fermion condensate,\footnote{%
  See, however, the discussion of the sign of $\tc_{00}$ in Sec.~\ref{c1lim}.
}
which, at the leading order, depends on the mass anomalous dimension
at the sill of the conformal window, as well as on two other
low-energy constants.

The GMOR-like relation for the dilatonic meson gives rise to a definition
of the expectation value $\svev{F^2}$ within the low-energy theory.
The definition of this expectation value in the microscopic theory is
ambiguous,
because of the power-divergent mixing of $F^2$ with the identity operator;
we leave the implications of a definition through the low-energy theory as
an open question.
For related discussions in the context of a lattice definition of
the energy-momentum tensor, see, \eg, Refs.~\cite{GP,dDZ}.
The behavior of $\svev{F^2}$ also plays a role in the
finite-temperature phase transition, see for example Refs.~\cite{BG,SF}.

Our work was motivated by lattice studies of asymptotically free models
whose spectrum contains a light flavor-singlet scalar meson
in certain parts of the bare-parameter phase diagram.
These models naturally provide a test bed for the effective field theory
framework we have developed.  The framework must be put to a test because,
unlike the standard chiral lagrangian, we cannot derive
the low-energy expansion from first principles; it is a consequence
of our main dynamical assumption, Eq.~(\ref{tbg}).

At the same time, the new low-energy framework provides a useful
analytic tool.  It is notoriously difficult to firmly
establish by numerical studies whether a particular model with
a slowly running coupling is chirally broken or infrared conformal.\footnote{%
  As an example, a recent study using domain-wall fermions suggests
  that the $SU(3)$ theory with $N_f=10$ fundamental fermions is
  infrared conformal \cite{Nf10},
  whereas another recent study using staggered fermions
  suggests that the $N_f=12$ theory could be chirally broken \cite{Nf12}.
  The results of these two studies are in conflict.
}
In a theory with an IRFP, the infrared physics satisfies
hyperscaling relations.  By contrast, the effective low-energy theory
developed here
predicts a distinctively different dependence of the light spectrum
on the fermion mass $m$, already at leading order.
Moreover, in an infrared-conformal theory, the inverse linear size
of the system is always a relevant operator, whereas in the effective
low-energy theory finite volume corrections occur only at the
next-to-leading order.  These qualitative differences
may help diagnose whether a specific model is infrared conformal or
chirally broken.

The low-energy framework would be best tested by exploring
many theories with fermions in the fundamental representation
with varying numbers of colors and flavors.
Since cost considerations may make such an extensive study difficult,
we have briefly commented on tests that can be
carried out within a given model with fixed values of $N_c$ and $N_f$.

Finally, while the low-energy framework is not applicable
to theories with higher-representation fermions, we noted that
in this case one may attempt a similar expansion, in which,
in place of the two small parameters $1/N$ and $n_f-n_f^*$,
there is a single small parameter $N_f-N_f^*$.  The meaning of $N_f^*$ is
that, by varying the number of flavors $N_f$ continuously,
the resulting (in general non-local) statistical mechanics system
is assumed to become infrared conformal at some non-integer value $N_f=N_f^*$.
The physical model where $N_f$ is the largest integer smaller than $N_f^*$
is chirally broken, and
might be amenable to an an expansion in $m/\L_{IR}$ and $N_f-N_f^*$
if the latter happens to be numerically small.

%%%%%%%%%%%%%%%%%%%%%%%%%%%
\vspace{3ex}
%\newpage
\noindent {\bf Acknowledgments}
\vspace{3ex}
%%%%%%%%%%%%%%%%%%%%%%%%%%%

We thank Adam Schwimmer and Ben Svetitsky for discussions.
We also thank the Centro de Ciencias de Benasque Pedro Pascual,
where this work was started, for hospitality.
This material is based upon work supported by the U.S. Department of
Energy, Office of Science, Office of High Energy Physics, under Award
Number DE-FG03-92ER40711.
YS is supported by the Israel Science Foundation
under grant no.~449/13.

%%%%%%%%%%%%%%%%%%%%%%%%%%%
%\newpage
\appendix
\section{\label{VZlim} Two-loop beta function in the Veneziano limit}
%%%%%%%%%%%%%%%%%%%%%%%%%%%
In terms of an 't Hooft coupling with a slightly different normalization
(compare Eq.~(\ref{tHooft}))
\begin{equation}
  \ta = g^2 N_c/(16 \p^2) \ ,
\label{g2N}
\end{equation}
the two-loop beta function for $N_f$ Dirac fermions in the
fundamental representation takes the form
\begin{subequations}
\label{betalph}
\begin{eqnarray}
  \b(\ta) \equiv \frac{\partial \ta}{\partial\log\mu}
  &=& -b_1\ta^2 -b_2\ta^3 \ ,
\label{betalpha}\\
  b_1 &=& 11/3 - 2n_f/3 \ ,
\label{betalphb}\\
  b_2 &=& 34/3 - n_f (13/3 - 1/N_c^2) \ ,
\label{betalphc}
\end{eqnarray}
\end{subequations}
where $n_f$ was defined in Eq.~(\ref{nf}).  In the Veneziano limit
we may drop the last term on the right-hand side of Eq.~(\ref{betalphc}).
In the two-loop approximation,
an IRFP exists for $n_f$ values in the following range
\begin{equation}
  2.61 \simeq 34/13 < n_f<5.5 \ .
\label{bzfp}
\end{equation}
For $n_f>5.5$, asymptotic freedom is lost.  The two-loop prediction
for the value of the IRFP is
\begin{equation}
  \ta_*(n_f) = (11-2 n_f)/(13 n_f - 34) \ .
\label{alphastar}
\end{equation}

Of course, Eq.~(\ref{alphastar}) is not reliable when $n_f$
is close to 34/13.  Moreover, the two-loop prediction is oblivious to
the possible onset of spontaneous chiral symmetry breaking.
As discussed in Sec.~\ref{conj}, for walking theories
we may combine the gap equation and the two-loop beta function to
postulate that chiral symmetry breaking sets in
when the running coupling reaches $\a_c$, provided that $\a_c<\a_*$.
The critical value of the rescaled coupling is
$\ta_c=\a_c/(4\p)=1/6$.  Substituting this value into Eq.~(\ref{betalph})
provides a model estimate for Eq.~(\ref{tbg}) (dividing out $[F^2]$), given by
\begin{equation}
  \b(\ta_c) = \b(1/6) = \frac{25}{648}\,(n_f-4) \ ,
\label{modeltbg}
\end{equation}
which corresponds to $n_f^*=4$ and $\eta=1$.  In other words,
the estimated range of the conformal window in the Veneziano limit is
$n_f^*=4 < n_f < 5.5$.

%%%%%%%%%%%%%%%%%%%%%%%%%%%
%\newpage
\section{\label{inductiveV} Details of the inductive proof}
%%%%%%%%%%%%%%%%%%%%%%%%%%%
In this appendix we complete the proof of Eq.~(\ref{cIkn})
by considering the remaining leading-order potentials
$V_M$, $V_\p$ and $V_\t$.  We will also briefly discuss the potentials
encountered at the next-to-leading order and beyond.

We start with $V_M$.  In order to probe this potential we add
to the $n$ differentiations with respect to $\s$ one more differentiation
with respect to the scalar source $\cs$, all at the same spacetime point.
We obtain
\begin{equation}
  \GEFT_{M,n} = \svev{\O_{M,n}(x) + \O_{M,0}(x)\,
  (\O_{d,n}(x)+\O_{\p,n}(x)+\O_{\t,n}(x)) } \ ,
\label{GeftVM}
\end{equation}
where
\begin{equation}
  \O_{M,n} = \frac{1}{2N_f} \, V_M^{(n)}(\t)\, e^{y\t} \tr (\S+\S^\dagger) \ .
\label{nMvertex}
\end{equation}
Note that $\O_{M,0}=\O_\cs/N_f$.
The additional normalization factor of $1/N_f$ relative to Eq.~(\ref{Ocs})
is introduced in order that $\GEFT_{M,n}$ will be $O(1)$ in large-$N$ counting.
The $\O_{M,n}$ term in Eq.~(\ref{GeftVM}) is obtained
by applying all $\s$-derivatives to $V_M$.
The other term is obtained when only the $\cs$ differentiation
is applied to $\tcl_m$, while the $\s$-derivatives are applied
to some other term in the lagrangian.

For the parameter range~(\ref{epsdelta}) we may neglect all contributions
of $\O_{\p,n}$ and $\O_{\t,n}$ relative to the leading contribution of
$\O_{d,n}$, for the same reasons as in Sec.~\ref{match}.  As for $\O_{d,n}$ itself,
at this point we have already established that $V_d$ satisfies
the power-counting hierarchy~(\ref{cIkn}).  Therefore, the contribution
coming from $\O_{M,n}$ must also be $O(\d^n)$.  It is easily seen
that this requires $c_{M,n}=O(\d^n)$.  The proof that $c_{M,k}=O(\d^n)$
for $k>n$ works in the same way as in Sec.~\ref{match}, by adding
$k-n$ external $\t$ legs.

The next potential to consider is $V_\p$.  Probing it requires
an operator that can serve as an interpolating field for a pion state,
which may be obtained by differentiating with respect to the pseudoscalar
source $\cp$.  Keeping the (suppressed) flavor indices open,
and using the same normalization factor as in Eq.~(\ref{Ocs}), we have
\begin{eqnarray}
  \O_\cp &=& -\frac{1}{2\hf_\p^2 \hB_\p} \frac{\partial \tcl}{\partial \cp}
\label{Ocp}\\
  &=& \frac{i}{4} \, V_M(\t)\, e^{y\t} (\S^\dagger-\S)
\NON
  &=& V_M(\t)\, e^{y\t} ( \p/\hf_\p + \cdots ) \ .
\nonumber
\end{eqnarray}
The basic correlation function we now consider involves the $n$
differentiations with respect to $\s(x)$, plus two differentiations
with respect to $\cp$ at spacetime points which are far apart.
We thus start by considering $\tGEFT_{\p,n}(q_1,q_2)$ which is
the Fourier transform of
\begin{equation}
  \GEFT_{\p,n}(x,z_1,z_2)
  = \svev{\Big(\O_{d,n}(x)+\O_{\p,n}(x)+\O_{\t,n}(x)\Big)\,
    \O_\cp(z_1)\, \O_\cp(z_2)} \ .
\label{Gnpitau}
\end{equation}
Once again we may ignore $\O_{d,n}$,
for which we have already proved Eq.~(\ref{cIkn}).
The only way to construct a tree diagram out of the remaining terms
is to connect $\O_{\p,n}$ by two pion lines to $\O_\cp(z_1)$ and $\O_\cp(z_2)$
(with a spacetime derivative acting on each line, see Eq.~(\ref{nvertexp})).
Since Weinberg's theorem ensures that loop diagrams are subleading,
we conclude that the leading contribution to $\tGEFT_{\p,n}(q_1,q_2)$
is proportional to $c_{\p,n}$, implying that $c_{\p,n}$ satisfies Eq.~(\ref{cIkn}).
As usual, in order to probe $c_{\p,k}$ for $k>n$ we add $k-n$ additional
$\t$ legs generated by insertions of $\tO_\cs$.

Finally we consider $V_\t$.  The first correlation function we consider
is $\tGEFT_{d,n,2}(p_1,p_2)$ (see Eq.~(\ref{GeftVdk})).  The contribution
of $\O_{\p,n}$ can be dropped since it will necessarily involve loops.
Based on its momentum dependence,
we isolate the tree diagram in which $\O_{d,n}+\O_{\t,n}$ is connected
by a $\t$ line to each of the two insertions of $\tO_\cs$.  After amputation,
the leading-order result is proportional to
\begin{equation}
  8\hB_\t c_{d,n} + p_1\cdot p_2\, c_{\t,n} \ ,
\label{Vtaun}
\end{equation}
(where, in the case of $\O_{d,n}$,
the two powers of $\t$ are obtained by expanding $e^{4\t}$ to second order).
We may now use the momentum dependence once more
to isolate the $c_{\t,n}$ term, which must therefore satisfy Eq.~(\ref{cIkn}).
The generalization to $c_{\t,k}$ for $k>n$ is similar.

\medskip

Invariant potentials $V_I(\t-\s)$ occur not only at the leading order,
but also at all higher orders.  Every higher-order term that arises
from what we have referred to in Sec.~\ref{lagnlo}
as the chiral expansion will be multiplied by its own
invariant potential.  The proof of Eq.~(\ref{cIkn}) generalizes to all
these potentials, too.

A concrete example illustrates how this works.
Consider once again $\tGEFT_{d,n,2}(p_1,p_2)$, which we have used
in order to study the $n$-th order term in the potential $V_\t$
occurring in the leading-order lagrangian.
We have isolated the relevant (amputated) tree diagram,
and calculated it up to second order in the external momenta, see Eq.~(\ref{Vtaun}).
If we extend the calculation of this correlation function to $O(p^4)$,
there will be two new kinds of contributions.
First, there will be one-loop diagrams involving
the LO vertices, and thus the LO potentials.
An additional $O(p^4)$ contribution will come from
the NLO operator $Q^\t_2$ of Eq.~(\ref{Q2}), which serves as a counterterm.
Since we already know that the LO potentials satisfy Eq.~(\ref{cIkn}),
we can now isolate the contribution of $Q^\t_2$, and thus, prove
that the invariant potential by which $Q^\t_2$ is multiplied must satisfy
Eq.~(\ref{cIkn}) too.  When we consider more external $\t$ legs, the other two
NLO operators in Eq.~(\ref{Q}) will in general contribute as well.
But, the fact that these operators are linearly independent means
that we must always be able to isolate the contribution of each one
of them, and thus, enforce the power-counting hierarchy on the corresponding
invariant potential.

%%%%%%%%%%%%%%%%%%%%%%%%%%%
%\newpage
\section{\label{diltc} Transformation properties of the potentials $V_I(\t)$}
%%%%%%%%%%%%%%%%%%%%%%%%%%%
Here we show that the invariance of a potential $V(\t-\s)$
under the dilatation transformation of Eqs.~(\ref{baredild}) and~(\ref{scleffb})
can be maintained when we set $\s=0$.  The invariance of $V(\t)$,
or, more precisely,
the transformation rule $V(\t(x)) \to V(\t(\l x))$, is achieved by promoting
the power-series coefficients to global spurions,
whose transformation rules will be derived below.
Dropping the coordinates dependence,
we start by rearranging the expansion of $V(\t-\s)$ in powers of $\t$,
\begin{eqnarray}
  V &=& \sum_{n=0}^\infty \frac{c_n}{n!}\, (\t-\s)^n
\label{expandVgena}\\
  &=& \sum_{n=0}^\infty \sum_{k=0}^n \frac{c_n}{k!(n-k)!}\, \t^k (-\s)^{n-k}
\NON
  &=& \sum_{k=0}^\infty  \frac{\hatc_k(\s)}{k!} \, \t^k \ ,
\nonumber
\end{eqnarray}
where we have introduced
\begin{equation}
  \hatc_k(\s) = \sum_{m=0}^\infty \frac{c_{m+k}}{m!} \, (-\s)^m \ .
\label{expandVgend}
\end{equation}
Note that $c_n=\hatc_n(0)$.
Under the transformation $\s \to \s + \D$ (where $\D=\log\l$)
we have, trivially, $\hatc_n(\s) \to \hatc_n(\s+\D)$.
The spurionic transformation of $c_n$ is derived by interpreting it
as $\hatc_n(0)$,
\begin{equation}
  c_n = \hatc_n(0) \to  \sum_{m=0}^\infty \frac{c_{m+n}}{m!} \, (-\D)^m \ .
\label{transchat}
\end{equation}

The shift by $\D=\log\l$ is parametrically $O(1)$, \ie, it does not
mix different orders in the expansion of the $c_n$'s in powers of
$n_f-n_f^*$, see Eq.~(\ref{expandcn}).
Noting that the double-expansion coefficients $\tc_{nk}$ vanish for $k<n$,
we find that their spurionic transformation rule is
\begin{equation}
  \tc_{nk} \to  \sum_{m=0}^{k-n} \frac{\tc_{m+n,k}}{m!} \, (-\D)^m \ ,
  \qquad k\ge n \ ,
\label{transctilde}
\end{equation}
which now involves only a finite number of terms on the right-hand side.

%%%%%%%%%%%%%%%%%%%%%%%%%%%
%\newpage
\section{\label{SLO} Derivation of the leading-order dilatation current}
%%%%%%%%%%%%%%%%%%%%%%%%%%%
In this appendix we derive the expression for the dilatation current
to leading order in the effective theory.
For more generality, we start from Eq.~(\ref{LeffL}), where we set
$\s(x)=0$ and $\c(x)=m$.  The infinitesimal transformations are
(compare Eqs.~(\ref{baredil}) and~(\ref{scleff}))
\begin{subequations}
\label{dfield}
\begin{eqnarray}
  \d\t &=& \th ( x_\n \partial_\n \t +1 ) \ ,
\label{dfielda}\\
  \d\S &=& \th x_\n \partial_\n \S \ .
\label{dfieldb}
\end{eqnarray}
\end{subequations}
Promoting $\th$ to a local parameter we find
(allowing for integration by parts since
we may choose $\th(x)$ to vanish at infinity)
\begin{subequations}
\label{dL}
\begin{eqnarray}
  \d_\th \tcl_\t &=& - \th \partial_\m \left[
    x_\n \left( T_{\m\n}^\t + \frac{1}{3} K_{\m\n} \right) \right]
  +\th \frac{f_\t^2}{2}\,  e^{2\t} (\partial_\m \t)^2 V'_\t \ ,
\label{dLa}\\
  \d_\th \tcl_\p &=& - \th \partial_\m ( x_\n T_{\m\n}^\p )
  +\th \frac{f_\p^2}{4}\,  e^{2\t}
            \tr[(\partial_\m \S^\dagger) (\partial_\m \S)] V'_\p \ ,
\label{dLb}\\
  \d_\th \tcl_d &=& \th f_\t^2 B_\t [ \partial_\n ( x_\n e^{4\t} V_d )
    + e^{4\t} V'_d ] \ ,
\label{dLc}\\
  \d_\th \tcl_m &=& -\frac{\th f_\p^2 B_\p m}{2}\, \Big[
    \partial_\n \Big( x_\n e^{y\t} V_M \tr(\S+\S^\dagger) \Big)
\label{dLd}\\
  && + e^{y\t} ((y-4)V_M+V'_M) \tr(\S+\S^\dagger) \Big] \ ,
\nonumber
\end{eqnarray}
\end{subequations}
where
\begin{subequations}
\label{Tmn}
\begin{eqnarray}
  T_{\m\n}^\t &=& \frac{f_\t^2 e^{2\t} V_\t}{2}
  \left( 2(\partial_\m \t) (\partial_\n \t)
         - \d_{\m\n} (\partial_\r \t)^2 \right) \ ,
\label{Tmna}\\
  T_{\m\n}^\p &=& \frac{f_\p^2 e^{2\t} V_\p}{4}
  \tr[(\partial_\m \S^\dagger) (\partial_\n \S)
      + (\partial_\n \S^\dagger) (\partial_\m \S)
      - \d_{\m\n} (\partial_\r \S^\dagger) (\partial_\r \S)] \ , \hspace{5ex}
\label{Tmnb}\\
  K_{\m\n} &=& f_\t^2 ( \d_{\m\n} \bo - \partial_\m \partial_\n ) \U \ ,
\label{Tmnc}
\end{eqnarray}
\end{subequations}
and $\U(\t)$ is defined by
\begin{equation}
  \frac{\partial \U}{\partial \t} = e^{2\t} V_\t \ .
\label{UeV}
\end{equation}
Note that $K_{\m\n}$ is transversal by construction.
From these results we read off the expression
for the dilatation current,
\begin{eqnarray}
  S_\m &=& x_\n \Th_{\m\n} \ ,
\label{LOcrnt}\\
  \Th_{\m\n} &=& T_{\m\n}^\t + T_{\m\n}^\p -\d_{\m\n} (\tcl_d + \tcl_m)
  + \frac{1}{3}\,  K_{\m\n}
\label{LOimp}\\
  &=& T_{\m\n} + \frac{1}{3}\,  K_{\m\n} \ ,
\nonumber
\end{eqnarray}
where $T_{\m\n}$ and $\Th_{\m\n}$ are the canonical and improved
energy-momentum tensors, respectively.

In order to obtain the leading-order expression for $S_\m$,
we substitute in the above results
$V_M=V_\p=V_\t=1$, and use Eq.~(\ref{Vdnf}) for $V_d$.
The solution of Eq.~(\ref{UeV}) is then $\U =  e^{2\t}/2$,
and the final expression for $K_{\m\n}$ may be found in Eq.~(\ref{Kmn}).

%%%%%%%%%%%%%%%%%%%%%%%%%%%
%\newpage
\section{\label{resumVeff} Resummation of one-loop diagrams}
%%%%%%%%%%%%%%%%%%%%%%%%%%%
Here we calculate the one-loop effective potential by resumming diagrams.
As in Sec.~\ref{NLO} we set $\tc_{00}=0$, but keep $\tc_{01}$ free.
The quantum field $\tilt$ is defined by Eq.~(\ref{cnnprop}).
We encounter two types of vertices.
The first possibility is that the vertex comes from
the kinetic term~(\ref{Lt}), in which case
both derivatives must act on the internal lines.
The rest of this vertex is given by
\begin{equation}
  \ca = e^{2\tilt/\hf_\t} - 1 \ .
\label{extV}
\end{equation}
The other type of vertex comes from expanding $\cl_d$ or $\cl_m$,
and, apart from the two internal lines, it is given by
\begin{equation}
  \cb = e^{-2v} (\cv''(v+\tilt/\hf_\t)-\cv''(v)) \ ,
\label{vrtxV}
\end{equation}
where $\cv$ was defined in Eq.~(\ref{taupot}).
The one-loop diagram with $n$ insertions of $\ca$ and $\ell$ insertions
of $\cb$ can be obtained as
\begin{equation}
  \G_{n,\ell} = \frac{1}{2}
  \left(\frac{\ca^n}{n!} \frac{\partial^n }{\partial \l^n}\right)
  \left(\frac{\cb^\ell}{\ell!}
  \frac{\partial^\ell }{\partial (m_\t^2)^\ell}\right)
  I_0\, \Bigg|_{\l=1} \ ,
\label{Gnl}
\end{equation}
where in $d\ne 4$ dimensions,
\begin{equation}
  I_0 = \G(-d/2) \left(\frac{m_\t^2}{4\p\l}\right)^{d/2} \ .
\label{I0}
\end{equation}
Summing over $n$ and $\ell$, and using $m_\t^2=e^{-2v} \cv''(v)$, we find
\begin{eqnarray}
  V_1 &=& -\sum_{n,\ell=0}^\infty \G_{n,\ell}
\label{V1diag}\\
  &=& -\frac{1}{2}\,
  \exp\left( \cb \frac{\partial }{\partial (m_\t^2)} \right)
  \exp\left( \ca \frac{\partial }{\partial \l} \right) I_0\, \Bigg|_{\l=1}
\NON
  &=& -\frac{\G(-d/2)}{2}
  \left(\frac{e^{-2(v+\tilt/\hf_\t)} \cv''(v+\tilt/\hf_\t)}{4\p}\right)^{d/2}
  \rule{0ex}{5ex}
\NON
  &=& -\frac{\G(-d/2)}{2}
  \left(\frac{e^{-2\t} \cv''(\t)}{4\p}\right)^{d/2} \ ,
\nonumber
\end{eqnarray}
in agreement with Eq.~(\ref{V1}).
The role of the first exponential on the second line
is to replace $m_\t^2$ in Eq.~(\ref{I0}) by $\cb+m_\t^2=e^{-2v} \cv''(\t)$,
while the role of the second is to replace $\l$ by $e^{-2v}e^{2\t}$.
The factors of $e^{2v}$ then cancel out, leading to the final result
which depends only on the full $\t(x)$ field.
The (minimally subtracted) renormalized effective potential
may be obtained from Eq.~(\ref{V1d}) by discarding the pole part,
and substituting $u(x)\to\t(x)$.

%\newpage
\vspace{5ex}
%%%%%%%%%%%%%%%%%%%%%%%%%%%

\end{document}